\documentclass[prd,aps,nofootinbib]{revtex4} 

\usepackage{latexsym, graphicx}
\usepackage{color, hyperref}

\begin{document}

\title{Seeing through a nearly black star}
\author{Shih-Yuin Lin}
\email{sylin@cc.ncue.edu.tw}
\affiliation{Department of Physics, National Changhua University of Education, Changhua 50007, Taiwan}

\date{\today}

\begin{abstract}
A nearly black, gravitationally intense star of semi-transparent, spherical, massive shell containing a few pointlike light sources inside would be perceived not like a three-dimensional ball for a localized observer outside the shell in terms of the affine or binocular distance. As the radius of the spherical shell approaches the Schwarzschild radius, the perceived distance between the front and rear surfaces of the shell would go to zero, while the images of most of the interior emitters would squeeze around the shell surfaces in terms of the affine or binocular distance. So, the Schwarzschild black hole formed from the star would be thought of as a two-dimensional membrane for the observers who can only measure the binocular distance and/or affine distance. However, the depth information of a point source inside the nearly black star can still be resolved in terms of the radar or luminosity distance, which needs the knowledge about the radar signals or standard candles sent in earlier by the observer outside the star. This suggests that at late times of gravitational collapse the area law of the entropy would dominate over the volume law for outside observers due to the loss of the knowledge about the ingoing probes earlier. 
\end{abstract}

\maketitle

\section{Introduction}

A field $\phi_{\bf x}(t)$ can be considered as a collection of oscillators, each oscillator is labeled with a space point ${\bf x}$ and evolving in time $t$, and each is interacting with the oscillators labeled as its spatial neighbors (and perhaps with itself, too) \cite{Go80, Lin12}. Since the entropy of a simple system in statistical physics is proportional to the dimension of the phase space, or the degrees of freedom of the system \cite{Hu87}, one may expect that the entropy of a field in a spatial region would be proportional to the number of oscillators, namely, the number of space points in that region, or the volume of that region divided by a constant specific volume.
It is therefore curious that the entropy of a black hole is proportional to the area rather than the volume of the black hole in classical general relativity \cite{BCH73}.

Much effort has been made in various perspectives to understand how the entropy of a black hole should be proportional to its area rather than its volume. The major direction is to count the degeneracy of microscopic states to each single macroscopic state of a stationary black hole \cite{Be73, Be75}. The leading-order results of such counting in string theory \cite{Co17} and loop quantum gravity \cite{Pe17} can match the area law of black hole entropy in many cases. However, the calculations in this direction are usually started with an eternal black hole whose horizon has been in existence. Having those interesting results, one may still ask the following: is there a continuous transition of entropy scaling from the volume law for a collapsing star to the area law for the black hole formed from that star? 

A sound answer comes from the study of the systems with negative specific heat due to the presence of long-range attractive forces \cite{Th70}. When the long-range interaction is gradually turned on, the thermodynamic entropy of the system can smoothly switch from extensive to non-extensive properties \cite{Op03}.
For a spherically symmetric star, as the star radius goes from a large number down to the Schwarzschild radius and the gravitational effect inside the star becomes more and more significant, the thermodynamic entropy of the star matter also shows a continuous transition from volume-scaling to area-scaling behaviors \cite{Op03, Op01, AYY19}. Along the transition, the local temperature grows while the entropy density decays inside the star. Eventually, the matter entropy of a nearly black star is mainly contributed by the pressure density of the star around the surface.
Nevertheless, a black hole's matter energy and pressure densities can be vanishing around its surface (i.e., the horizon), while its Bekenstein-Hawking entropy is large. There seems to be a gap between these results and the black hole entropy.

In this paper, we suggest an alternative thought.
Since the Hawking temperature associated with the Bekenstein-Hawking entropy in the equation of state originates from the vacuum fluctuations of quantum fields in the background geometry of a collapsing star \cite{Ha75}, the Bekenstein-Hawking entropy of a black hole should be proportional to the number of field degrees of freedom in the star.
Suppose no observer localized outside a star can communicate with the observers inside the star in a time-scale associated with their cutoffs, or during the period of interest in a model. Following the line in the opening paragraph, if {\it every} outside observer perceives the collection of space points in the star like a membrane rather than a ball, then for the outside world, the number of field degrees of freedom in the star should be considered as proportional to that area of the membrane rather than the volume of a ball. 
If all those two-way causally connected observers can in principle agree on a standard area for the membrane after they exchange and compare their results, the area-scaling number of field degrees of freedom should be considered as a relativistic invariance.

To show that such a situation would be possible, we are studying how the position information of the pointlike emitters distributed inside a spherical massive shell, which is about to form a black hole, would be perceived by an observer localized outside the star. We will see that, as the shell radius approaches the Schwarzschild radius, namely, the star becomes ``gravitationally intense" \cite{Sy66}, the interior of the shell would be perceived like a two-dimensional (2D) membrane rather than a three-dimensional (3D) ball in terms of the affine and binocular distances,\footnote{Similar ideas in the affine distance can be found in \cite{Hahttp}, while the direct images of the rear surfaces ($R$ in Figure \ref{plotDefdphi} for each observed light ray) of the collapsing star is not shown there.} so the outcome of each local measurement on the field amplitude or momentum in the star would be positioned on the 2D membrane. 
This implies that the Bekenstein-Hawking entropy is actually extensive and thermodynamically admissible in observational coordinates,
and one would not need to construct an alternative entropic functional to be used for thermodynamical issues \cite{Ts19}.

We will further see that the interior of the shell still looks like a 3D ball in terms of the radar or the luminosity distances whenever the nearly black star is not truly black. The key difference between these two kinds of distance measures suggests that, 
by including or ignoring the full knowledge about the probes and responses, namely, by keeping or dropping the autocorrelations of ingoing and outgoing signals, the field degrees of freedom inside the star can switch between a volume-scaling and an area-scaling quantity for a localized observer outside. At late times of gravitational collapse, the area law eventually dominates among the observations due to the growing difficulty of reconstructing those autocorrelations.

This paper is organized as follows. 
We focus our attention on a simple geometry produced by a semi-transparent spherical massive shell in Section \ref{SpheMassShell}, where we determine the affine distance from a point source of light inside or outside the spherical shell to an observer localized outside the star. In Section \ref{binoculardistances} we introduce the binocular distances perceived by the observer with baselines in two orthogonal directions to compare with the affine distances of the interior emitters. We further examine in Section \ref{lumin} if other measures of distance such as the radar distance and luminosity distance would give similar results. Then, we summarize and discuss our results in Section \ref{SumDisc}. A review on the null geodesic equations describing the light rays in a spherically symmetric spacetime is given in Appendix \ref{RevGeoEq}, and the angles of departure of the observable light rays sourced from the interior emitters are discussed in Appendix \ref{EscCone}. 

\section{Emitters in a spherical massive shell}
\label{SpheMassShell}

Consider a star of a spherical thin shell of radius $r_s$, total mass $M$, and centered at $C$ chosen as the origin of bookkeeper coordinates (\ref{SphereMetric}), containing a few pointlike light emitters of negligible masses 
to help the outside observers determining the locations of the events inside the spherical shell.\footnote{For more realistic cases, see \cite{Pe04, Op03, AYY19}}.
The mass of the star is concentrated on the shell of negligible thickness with a uniform surface density, and the shell is semi-transparent so that light rays can go through the shell and a localized observer at a fixed radius outside the star can see the front and the rear surfaces of the shell, like viewing a dusty hollow glass sphere, as well as those light emitters inside. Note that the light rays here would not only refer to some eikonal limit of realistic electromagnetic waves, which may be scattered by the interior matter of a star, but also the ideal light rays in relativistic physics to specify the causal structure \cite{HE73, Ha75}, or any messenger fields or particles weakly interacting with matter \cite{Ku65}. 

\begin{figure}
\includegraphics[width=8cm]{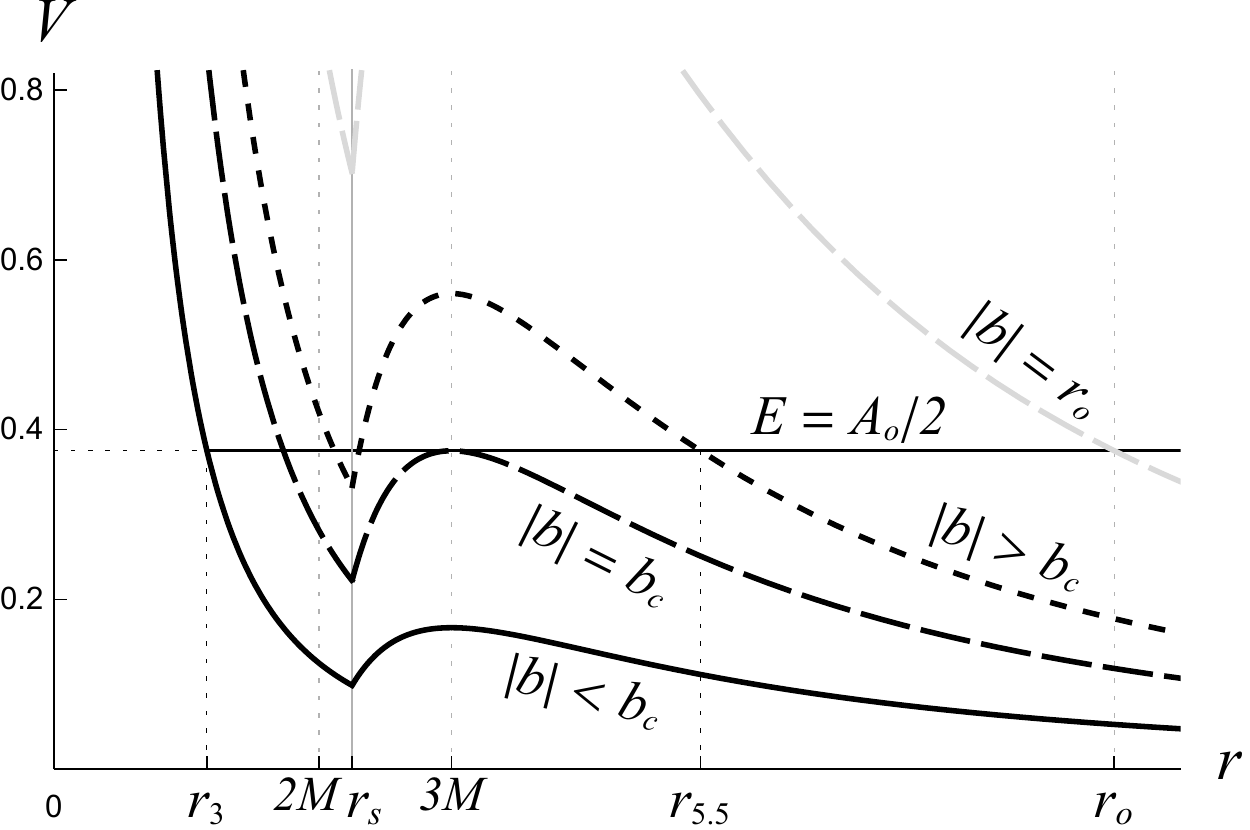}
\caption{The effective potential given in (\ref{Eshell}) 
with $b=3$ (solid curve), $b=b_c=4.5$ (dashed), $b=5.5$ (dotted), and $b=r_o$ (gray dashed). Local maxima occur at the photon sphere $r=3M$. Here, the total mass of the spherical shell is $M=1$; the shell and the observer are situated at $r=r_s=2.25M$ and $r_o=8 M$, respectively; and $b_c= 3\sqrt{3 A_o}M$ defined above Eq. (\ref{largestangle}). The horizontal line represents the effective total energy $E=A_o/2 =3/8$. 
The least values of the bookkeeper $r$ coordinate that the rays with $b=3$ ($< b_c$) and $5.5$ ($> b_c$) can reach are $r_{\rm min} = r_3 \approx 1.1547 M < 2M$ and $r_{5.5} \approx 4.8783 M > 3M$, respectively.}
\label{Vofr}
\end{figure}

Suppose the interior of the star is otherwise empty and the bending of light rays is purely due to the spacetime geometry (\ref{SphereMetric}), which has 
\begin{equation}
  A(r) = 1/B(r) = 1-\frac{2M}{r} \hspace{.5cm} {\rm for}\; r>r_s \label{Schsch}
\end{equation}
outside the star as the Schwarzschild metric by Birkhoff's theorem and 
\begin{equation}
  A(r) = A_s \equiv 1-\frac{2M}{r_s}, \hspace{.5cm} B(r)=1 \hspace{.5cm} {\rm for}\; r\le r_s
\end{equation}
inside as the Minkowski metric with a scaled time coordinate \cite{Is66, Po04, NMM16}. Note that $A_s\to 0$ as $r_s\to 2M$.
Then, Eq.(\ref{ETV}) can be written as $E=T+V$, where
\begin{eqnarray}
  T =  \left\{\begin{array}{l}
	        \dot{r}^2/2 \\
					A_s\dot{r}^2/2
       	 \end{array}\right. 
  \hspace{1cm}
  V =  \left\{\begin{array}{lll}
	        \frac{b^2}{2r^2}\left(1-\frac{2M}{r}\right) &\hspace{1cm}{\rm for} & r>r_s, \\
					\frac{b^2}{2r^2}A_s 
					  &\hspace{1cm}{\rm for} & r\le r_s ,
       	 \end{array}\right. \label{Eshell}
\end{eqnarray}
and the light rays of different $b$ have different effective potentials $V$ (Figure \ref{Vofr}).

When $r_s \ge 3M$, $V(r)$ is monotonically decreasing as $r$ increases. For a gravitationally intense star $2M < r_s < 3M$, as shown in Figure \ref{Vofr}, the Schwarzschild metric outside the shell forms a local maximum of $V$ at $r=3M$, where the photon sphere is located. The light rays of $E < V(3M)$, or $|b| > b_c$, where 
\begin{equation}
    b^{}_{c} = 3 M \sqrt{6E} 
		\label{bcrit}
\end{equation}
is defined by $E= 
V(3M)|_{b=b_c}$,
cannot cross the barrier of the effective potential. Such null geodesics that started from the inside of the spherical shell will be trapped in the photon sphere, and that started from the outside of the photon sphere cannot reach the surface of the spherical shell. 
Once the null geodesic can reach and enter the shell, the closest radius $r$ to the shell center it can possibly reach is
\begin{equation}
  r_{\rm min} \equiv \sqrt{\frac{b^2 A_s}{2E}} 
	\label{rminb}
\end{equation}
where $\dot{r}=0$ and so the effective kinetic energy $T=0$. For finite values of $E$ and $b$, one has $r_{\rm min} \to 0$ as $r_s\to 2M$. 

When considering the null geodesics passing through the localized observer at $r_o > r_s$, we insert $E=A_o/2$ from (\ref{EAo}) into (\ref{bcrit}) and (\ref{rminb}).

\subsection{Angle of arrival and affine distance}
\label{thetaadA}

\begin{figure}
\includegraphics[width=9cm]{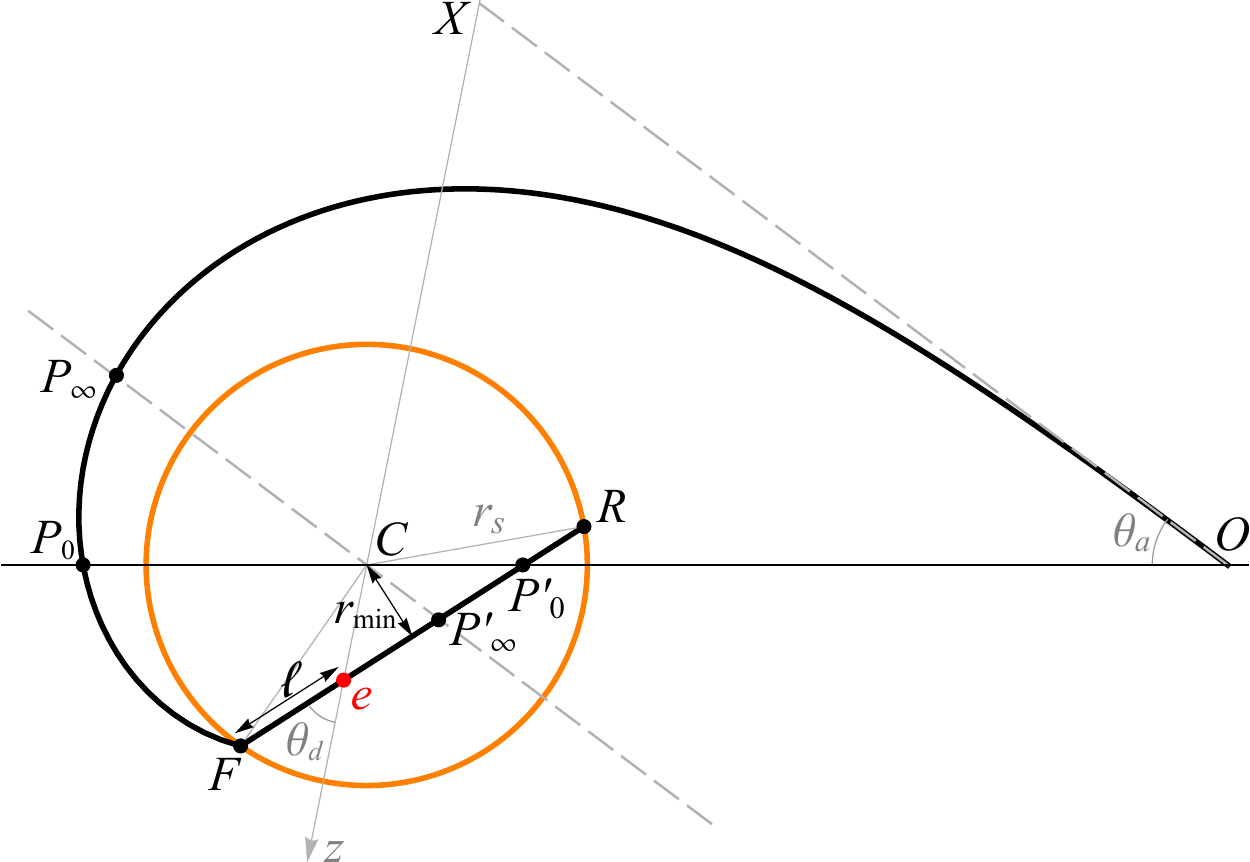}
\caption{
The black curve represents the light ray with $b\approx 4.36062$, and the spherical shell is colored in orange. 
Here, $M=1$, $r_s=2.05M$, and $r_o=8M$. The straight line joining the shell center $C$ and $P_\infty$ is parallel to the tangent line of the light ray around the observer $O$ (gray dashed lines).}
\label{plotDefdphi}
\end{figure}

\begin{figure}
\includegraphics[width=8cm]{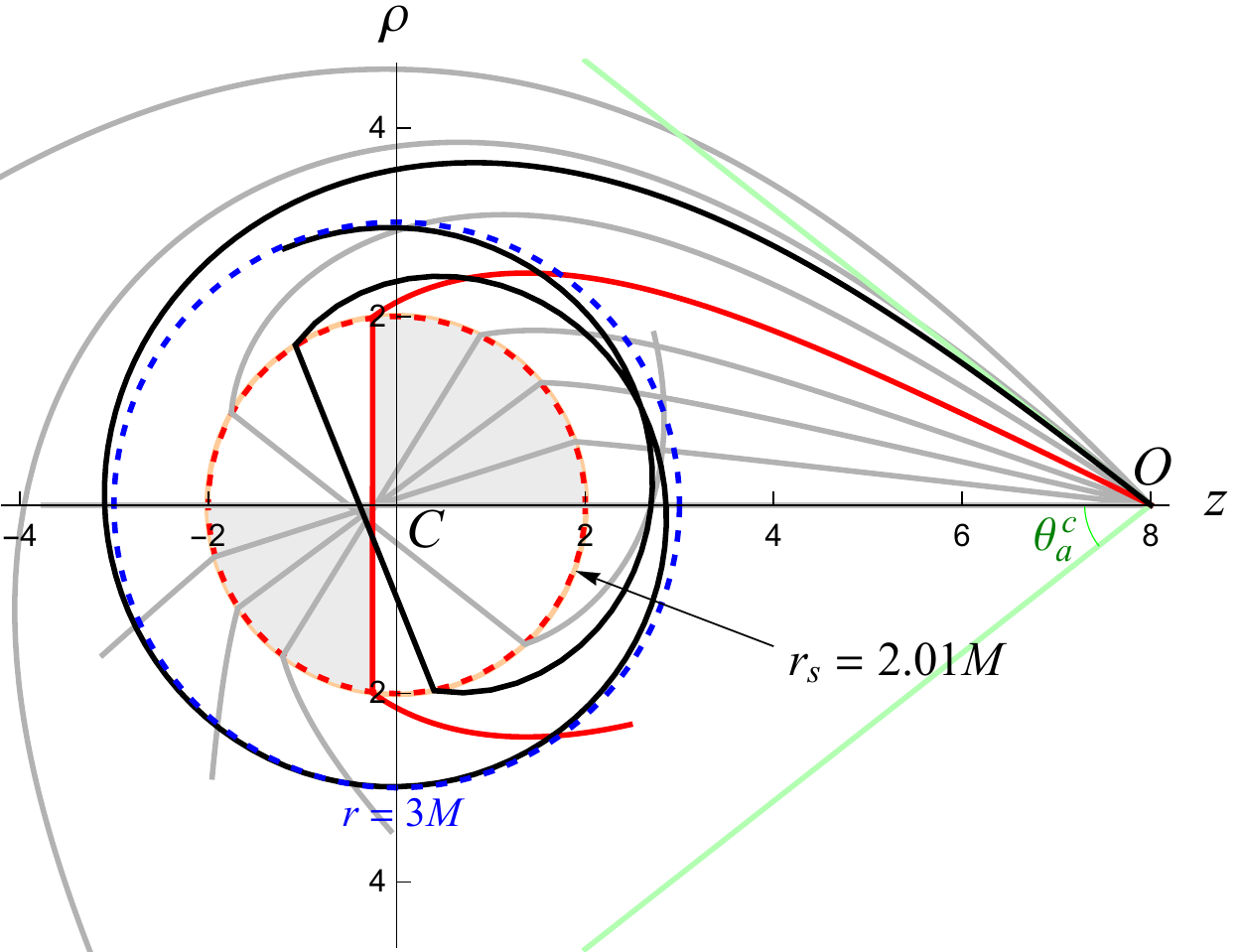}
\caption{The light rays (red, black, and gray curves) around a spherical shell (orange) and received by the localized observer $O$ are represented in bookkeeper coordinates $r$ and $\theta$, plotted as polar coordinates on a 2D plane (of some constant $\varphi$). Here, $M=1$, $r_o=8 M$, and $r_s=2.01M$. The photon sphere at $r=3M$ and the Schwarzschild radius $r=2M$ are represented in blue dotted and red dotted circles, respectively. The red ray has $\theta_a =\theta_1$, which is the minimum angle of arrival within which the localized observer can collect the signals from {\it all} the point sources inside the shell.} 
\label{theta1}
\end{figure}

The angle of arrival for a light ray received by the localized observer $O$ at $r=r_o$ is
\begin{equation}
  \theta_a = \left. \tan^{-1} \, \frac{r \dot{\theta}}{-\dot{r}}\right|_{r=r_o} = 
	    \tan^{-1} \frac{b/r_o}{\sqrt{A_o\left[1-(b/r_o)^2\right]}} \label{thetaab}
\end{equation}
in the $r\theta$ plane in bookkeeper coordinates, assuming that the observer is always facing to the shell center $C$ at the origin
(Figure \ref{plotDefdphi}). The perceived angle of arrival would then be
\begin{equation}
  \tilde{\theta}_a = \left. \tan^{-1} \, \frac{r_o \dot{\theta}}{-\dot{\tilde{r}}} \right|_{r=r_o}
	  = \tan^{-1} \frac{b/r_o}{\sqrt{1-(b/r_o)^2}} = \sin^{-1} \frac{b}{r_o}  \label{PAOA}
\end{equation}
in terms of the radar coordinates, $ds^2 = dr^2/A_o + r_o^2 d\theta^2 \equiv d\tilde{r}^2 +r_o^2 d\theta^2$ for $dt=d\varphi=0$, around the observer at $r=r_o$. Note that the value of $\theta_a$ here increases in the clockwise direction in Figures \ref{plotDefdphi} and \ref{theta1}, in contrast to other angles defined in this paper such as (\ref{thetad}) and (\ref{thetaio}).
The above $\theta_a$, $\tilde{\theta}_a$, and $b$ are allowed to have negative values. 
When some positive value of $b$ is associated with some value of $\theta_a$, then $\pi+\theta_a$ will also be associated with $b$, and $-b$ will be associated with $-\theta_a$ and $\pi-\theta_a$. 
Here, one should choose $b^2 \le r_o^2$ to make $\dot{r}(r_o)$ real and $V(r_o)\le A_o/2=E$ [(\ref{ETV}), (\ref{Veff}), (\ref{EAo}), and Figure \ref{Vofr}]. As $b\to \pm r_o$, one has $\theta_a, \tilde{\theta}_a\to \pm \pi/2$.

From (\ref{ETV}), the affine distance $d^{}_A$ of a null geodesic connecting an event or emitter $e$ at $(r_e,\theta_e)$ and the observer $O$ at $(r_o,\theta_o)$ is defined as the difference of the normalized affine parameter between $e$ and $O$, namely,
\begin{equation}
   d^{}_A = \int_O^e d\lambda = 
	  \int_O^e \sqrt{\frac{A(r)B(r)}{2(E-V(r))}}dr \label{affdist}
\end{equation}
where $E$ is given in (\ref{EAo}) and $\lambda\equiv 0$ at $O$.
The above integrand will be real if the connecting null geodesic exists classically. To make $d_A$ positive and monotonically increasing
as we move $e$ away from $O$ along the same null geodesic, one needs to take care of the upper and lower limits for the above $r$ integration, which should be done piecewise if $r(\lambda)$ is not single valued [cf. (\ref{thetaofr0}) and below].

For an observer outside the photon sphere ($r_o > 3M$, called a ``far observer" below), when
the event or emitter is outside the shell with the whole null geodesic going in the Schwarzschild geometry, one has
\begin{equation}
   d^{}_A = \left| \int_{r_e}^{r_o} \frac{r^2 dr}{\sqrt{A_o r^4 - b^2 r^2 + 2Mb^2 r}}\right|, \label{affdistOut}
\end{equation}
and when the event/emitter is inside the shell, one has
\begin{eqnarray}
    d^{}_A &=& \int^{F}_{e} \sqrt{\frac{A_s}{A_o- A_s b^2/r^2}}dr 
		+ \int_{r_s}^{r_o} \frac{r^2 dr}{\sqrt{A_o r^4 - b^2 r^2 + 2Mb^2 r}} \nonumber\\
		&=& \sqrt{\frac{A_s}{A_o}}\,\, \ell 
		+ \int_{r_s}^{r_o} \frac{r^2 dr}{\sqrt{A_o r^4 - b^2 r^2 + 2Mb^2 r}}, \label{affdistIn}
\end{eqnarray}
where $\ell$ is the depth of the emitter from the front surface of the shell, i.e., $|eF|$ in Figure \ref{plotDefdphi}. There, $F$ ($R$) represents the intersection of the null geodesic and the front (rear) surface of the shell with respect to the affine distance for the observer. In bookkeeper coordinates,
\begin{equation} 
  \ell \equiv \frac{L}{2} \mp \sqrt{r_e^2-r_{\rm min}^2} \label{defellL}
\end{equation}
with ``$-$" for $|eF|\le |eR|$ in Figure \ref{plotDefdphi}, ``$+$" for $|eF|>|eR|$, and the depth of the rear surfaces from the front surface $L\equiv 2\sqrt{r_s^2-r_{\rm min}^2}=|FR|$. Note that both $\ell$ and $L$ depend on $b$ due to the $b$ dependence in $r^{}_{\rm min}$ from (\ref{rminb}).

For an observer inside the photon sphere (called a ``near observer" below, only possible when the star is gravitational intense, namely, $2M < r_s < r_o < 3M$), the light rays of $b \in (b_c, r_o)$ will be trapped in the photon sphere and oscillate between two of the real solutions for $E=V(r)$, $r=r_{\rm min}$ given in (\ref{rminb}) and $r_{\rm max} \in [r_o, 3M)$. Then, $r$ in (\ref{affdist}) is not single valued in $\lambda$, and (\ref{affdist}) should be calculated more carefully.  

\begin{figure}
\includegraphics[width=6.2cm]{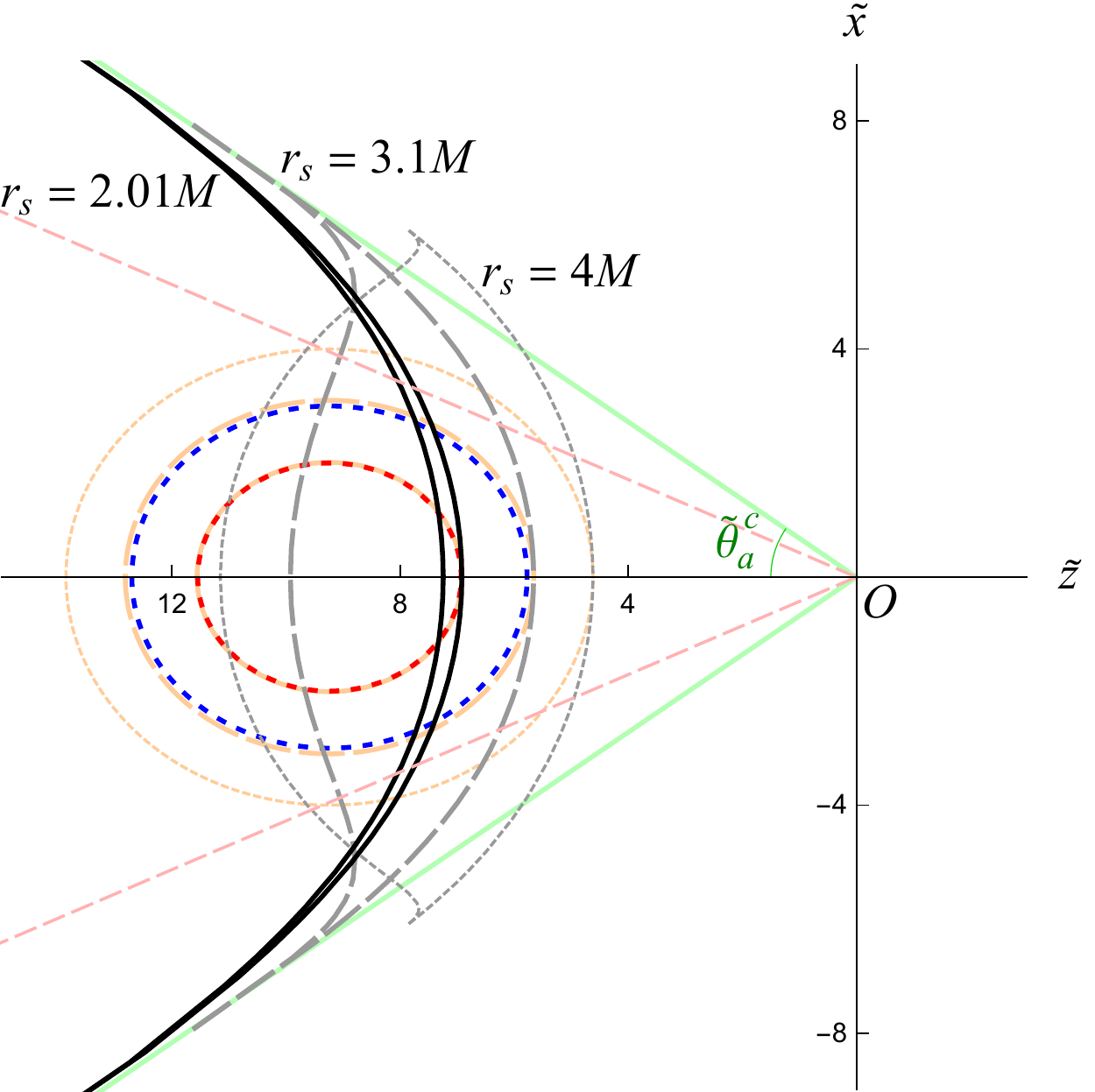}
\caption{The observed spherical shells of $r_s=4M$ (gray, short-dashed curve), $r_s=3.1M$ (gray, long-dashed), and $r_s = 2.01M$ (black) by a far observer at $r_o=8M$ in terms of the affine distance $d^{}_A$ and the angle of arrival $\tilde{\theta}_a$ in a constant-$\varphi$ plane in the observer's local frame. Here, $M=1$ and $(\tilde{z},\tilde{x}) \equiv (d^{}_A \cos \tilde{\theta}_a, d^{}_A \sin\tilde{\theta}_a)$. The black curves actually extend to infinity with the asymptotes in green. 
The red dashed lines represent $\tilde{\theta}_a=\pm\tilde{\theta}_1$, where $\tilde{\theta}_1 \approx 0.1286 \pi$ (with $b_1 \approx 0.25615$) for the case of $r_s = 2.01M$. 
For comparison, the spherical shells of different radii are represented in orange short-dashed, long-dashed, and solid curves, together with the photon sphere (blue dotted) and the Schwarzschild radius (red dotted) in the $r\theta$ plane of bookkeeper coordinates, with the $z$ coordinate scaled by $1/\sqrt{A_o}$. One can see that the affine distance between the front and rear surfaces of the spherical shell decreases as $r_s \to 2M$.}
\label{DisA}
\end{figure}

Around the shell surface $r=r_s$, since the effective potential $V$ in (\ref{Eshell}) is continuous and $E$ is a constant, the effective kinetic energy $T$ has to be continuous, too. This implies that $\dot{r}|_{r_s+\epsilon} = \sqrt{A_s} \dot{r} |_{r_s-\epsilon}$,  $\epsilon \to 0+$, and so for the light ray traveling for the same $\Delta r$ in the bookkeeper $r$ coordinate, the affine distance $\Delta\lambda$ just inside the shell is shrunk from the one just outside the shell by a factor $\sqrt{A_s} < 1$. 
As $r_s$ approaches the Schwarzschild radius $2M$ and so $\sqrt{A_s}\to 0$, the affine distance from the front to the rear surfaces of the shell for the observer, $\Delta\lambda=\sqrt{A_s/A_o}\,L$ from (\ref{affdistIn}), goes to zero at every angle of arrival, and the interior emitters of different depths between the two surfaces along the same light ray become more and more difficult to resolve for the observer. In other words, our nearly black spherical star ``perceived" by the localized observer is not a ball in terms of $(d^{}_A, \tilde{\theta}_a)$. Rather, it would look like a pancake or a contact lens by a far observer ($r_o>3M$), and will get into a membrane as its radius $r_s$ goes down to $2M$, as shown in Figure \ref{DisA}.
In the same plot, one can also see that earlier when $r_s$ was going down to the vicinity of the photon-sphere radius $3M$, the edge of the 3D map of the shell (in terms of the affine distance and the perceived angle of arrival) started to stretch backward along an asymptotic cone for the observer. When $2M < r_s <3M$, the affine distance of the edge of the 3D ``image" goes to infinity, and 
the angle of arrival of the boundary of the shell's image 
for the far observer localized at $r_o>3M$ corresponds to the rays coming exactly from the photon sphere $r=3M$ with $E = V(3M)$, namely,
\begin{equation}
  \tilde{\theta}_a^c =  \sin^{-1} \frac{b_c}{r_o} = \sin^{-1} \frac{3\sqrt{3 A_o}~M}{r_o} \label{largestangle}
\end{equation}
from (\ref{bcrit}) and (\ref{EAo}). For the far observer,
the received rays with $|b|>b^{}_c$ or $|\tilde{\theta}^{}_a| > \tilde{\theta}_a^c$ must have never been inside the spherical shell.

\begin{figure}
\includegraphics[width=7cm]{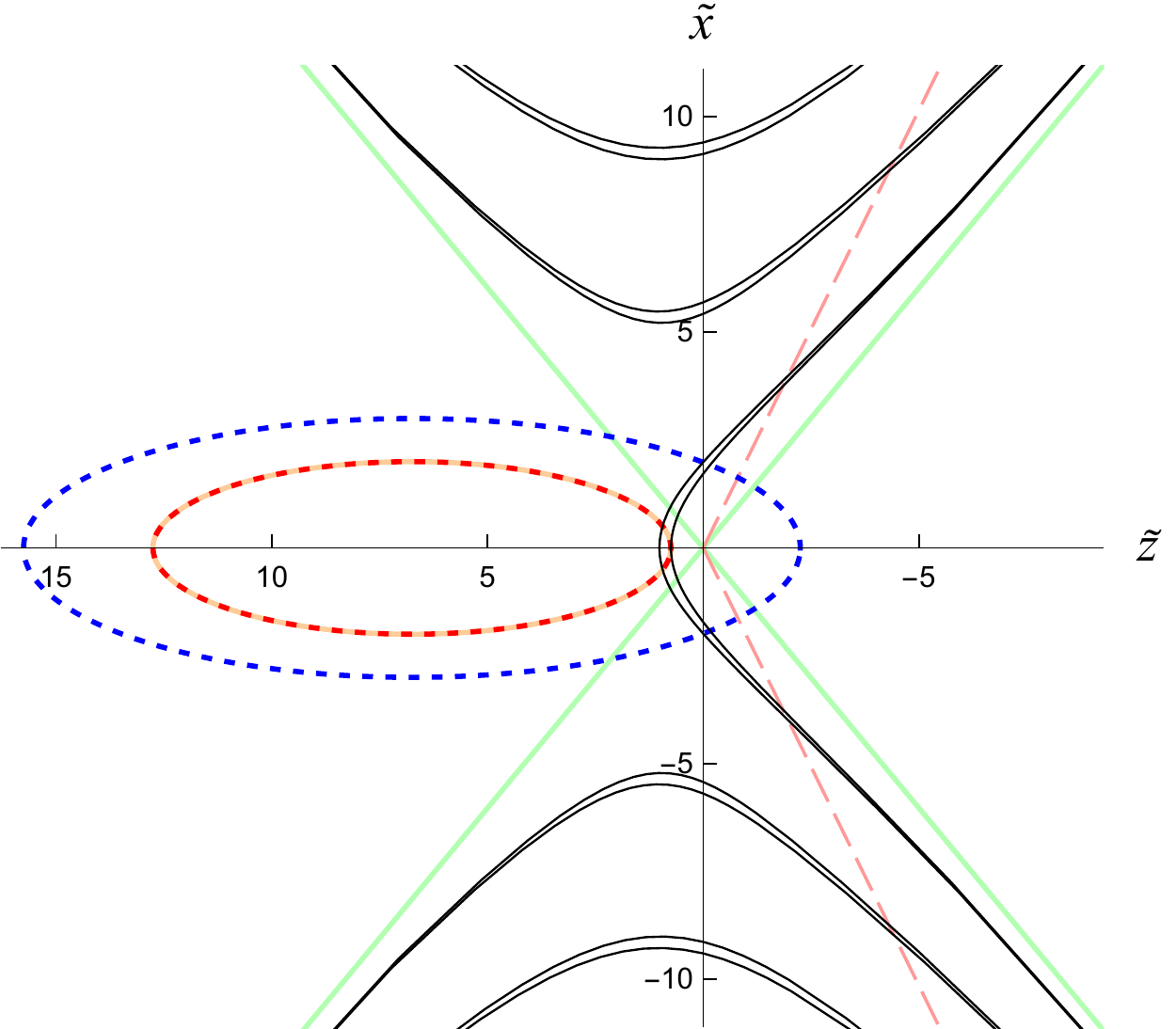}
\caption{The observed spherical shell of $r_s=2.001M$ (black curves) by a near observer at $r_o=2.25M$ inside the photon sphere in terms of the affine distance $d^{}_A$ and the angle of arrival $\tilde{\theta}_a$ in a constant-$\varphi$ plane in the observer's local frame. The green lines represent the angle of arrival $\tilde{\theta}_a^c$ and $\pi-\tilde{\theta}_a^c$, and the red-dashed lines represent $\tilde{a}=\pm\tilde{\theta}_1$, where $\tilde{\theta}_1 \approx 0.6457\pi > \pi/2$ (with $b_1\approx 2.0184$). The ellipses are the spherical shell (orange), the photon sphere (blue dotted) and the Schwarzschild radius (red dotted) in the $r\theta$ plane of bookkeeper coordinates, with the $z$ coordinate scaled by $1/\sqrt{A_o}$.}
\label{DisAinPS}
\end{figure}

For a near observer ($2M < r_s < r_o <3M$), the boundary of the star image would be perceived at the angle of arrival $\tilde{\theta}_a = \pi-\tilde{\theta}_a^c \ge \pi/2$. The images at sufficiently small angles of arrival $\tilde{\theta}_a \le \tilde{\theta}_a^c$ (which is one of the escape cones for a pointlike emitter situated at the same position of the localized observer $O$; see Appendix \ref{EscCone}) look similar to those in the above case for the far observers.
At larger angles of arrival $\tilde{\theta}_a\in (\tilde{\theta}_a^c, \pi-\tilde{\theta}_a^c)$, however, the images of the interior emitters are produced by the light rays trapped in the photon sphere and going into and out of the shell periodically. 
Those images would distribute in a series of thin layers of thickness $\sqrt{A_s/A_o}\,L$ (Figure \ref{DisAinPS}) separated by finite inter-layer gaps of the affine distances contributed by the sections of the light rays outside the shell.
At $\theta_a=\pi/2$, where $b$ reaches the maximum value $r_o$, the gap is minimal but still greater than the affine distance to the shell surface for the observer. 
Thus, it would not be difficult for the observer to distinguish the nearest image layer from the others in terms of the affine distance. 
As $r_s\to 2M$, while the affine distance to the thin layer where an image is positioned could be resolved by the observer, the affine depths of that image in its layer would not be resolvable. 

In short, for a localized observer outside a nearly black shell, either she is situated inside or outside the photon sphere of the shell, 
the information of depth of the interior emitters from the shell surface in terms of the affine distance will be lost as the shell is turning into a black hole.

\subsection{Direct images}
\label{dirimg}

In Figure \ref{theta1}, one can see that inside the nearly black star, 
as $b$ and so $\theta_a$ in (\ref{thetaab}) increase from zero, the sections of the light rays inside the shell roughly rotate about the shell center $C$ and scan the interior of the star on the same $r\theta$ plane. 
A further rotation of the whole $r\theta$ plane about the $z$ axis from $\varphi=0$ to $2\pi$ will make the shaded region in Figure \ref{theta1} pass through every interior point of the spherical shell at least once. 
Thus the red curve with the section inside the shell perpendicular to the $z$ axis in Figure \ref{theta1} represents the light ray with $\theta_a$ equal to the minimal angle of arrival $\theta_1$ within which the localized observer could possibly collect the signals emitted by {\it all} the point sources inside the shell (red dashed lines in Figure \ref{DisA}).

In the interval of the angle of arrival $|\theta_a|\le \theta_1$ for a far observer ($r_o > 3M$), most of the pointlike emitters inside the shell have single pointlike images. The only exceptions are those emitters located in the small region $-r_{\rm min}(b_1)<z<0$ with $b_1 \equiv b|_{\theta_a=\theta_1}= r_o \sin \tilde{\theta}_1$ around the $z$ axis (between the vertical red curve and the shell center $C$ in Figure \ref{theta1}), where the pointlike emitters right on the $z$ axis would produce small Einstein rings, and the ones slightly off the $z$ axis may produce double images asymmetric to the center of the image of the whole star (both kinds of these images will reduce to single pointlike images as $r_s\to 2M$). All of them are direct images. 

As one keeps increasing $|\theta_a|$ from $\theta_1$ toward $\theta_a^c$ (corresponding to $\tilde{\theta}_a^c$), the pointlike emitters in most of the region inside the shell start to be seen repeatedly as the first, second, and perhaps infinitely many higher-order indirect images in the picture of the whole star, if the whole setup is stable for an infinitely long period and the observer has an infinite angular resolution. However, the emitters in the core region of $r \le r_{\rm min}(b_1)$ around the shell center $C$ cannot be seen at any $|\theta_a| > \theta_1$ by the observer, since both $r_{\rm min}(b)$ and $\theta_a(b)$ of the light rays that the observer $O$ receives [Eqs. (\ref{rminb}) and (\ref{thetaab})] are increasing function of $|b|$, which implies that the light rays of $b > b_1$ or $|\theta_a|>\theta_1$ will always have $r_{\rm min}(b) > r_{\rm min}(b_1)$ and never reach the emitters in that core region of $r<r_{\rm min}(b_1)$ when traced back. 

Let the intersection of the red ray and the upper shell surface be $(r_s, \theta_{s1})$ in bookkeeper coordinates in Figure \ref{theta1}, where the observer $O$ is located at $(r_o,\theta_o)$. One can see that $\theta_{s1}-\theta_o$ is greater than $\pi/2$ in Figure \ref{theta1} because $r_{\rm min}$ is still significant there. As $r_s\to 2M$ and so $r_{\rm min}\to 0$, the polar angle $\theta_{s1}-\theta_o$ corresponding to $\theta_1$ will go to $\pi/2$, and the core region mentioned above will shrink to point $C$. 

For a near observer inside the photon sphere, $\theta_1$ can be defined in the same way as in Figure \ref{theta1} on the nearest image layer to the observer. As shown in Figure \ref{DisAinPS}, in this case, $\theta_1$ can be greater than $\theta_a^c$ and even greater than $\pi/2$ if $r_o$ is sufficiently close to $2M$. In spite of this, the observer would still be able to identify the direct images of all the interior emitters within $|\theta_a|\le \theta_1$ in the nearest image layer. 

Since $r_{\rm min}$ given in (\ref{rminb}) varies with different $b$, the direct images of two emitters inside the spherical shell may overlap (along the same ray of some value of $b$ to reach the observer), while their higher-order indirect images split (along two rays of different $b$), and vice versa. From this, an observer (or a group of observers situated at different angles in bookkeeper coordinates) outside the shell may be able to extract the relative positions of different interior emitters, including the relative depths of them. As $r_s\to 2M$, however, all the light rays from the interior emitters to the observer outside the shell will have their inside-shell sections almost going in the radial and opposite directions, namely, the straight-line extension of each inside-shell section will almost pass through the shell center $C$ in bookkeeper coordinates as all $r_{\rm min}(b)$ goes to zero in this limit (also see Appendix \ref{EscCone}). This implies that the images of all the interior emitters lying on the same diameter of the spherical shell would overlap in all orders and all layers when the shell is about to form a black hole, while the images of the interior emitters not on the same diameter would never overlap in view of all the observers localized outside the shell.
Then, it becomes impossible for those observers to extract the depth information of the interior emitters using the relative positions of their images in different orders or layers, from which only the emitters' angular positions can be determined.

\section{binocular distances}
\label{binoculardistances}

In previous section, we have learned that as a 3D spherical star is turning into a black hole, it would be perceived more and more like a 2D membrane by a localized observer outside the star in terms of the affine distance. One may argue that the affine distance is a mathematical construction, which is not measurable directly by physical means. Below, we examine whether the observation would be similar in terms of physical measurables such as the binocular distance.

To determine the binocular distance, our localized observer should set a baseline of non-zero length. We assume the baseline is infinitesimal to suppress the ambiguity. In a spherically symmetric spacetime, the binocular distance to a point source of light may be well determined using a baseline either parallel or perpendicular to the polar direction with respect to the $z$ axis joining the origin $C$ and the localized observer $O$ or the emitter $e$, though the values of the binocular distances determined by these two orthogonal baselines are different in general for the localized observer. Off these two directions, however, the binocular distance could not be determined straightforwardly, and the observer may need a more sophisticated way to obtain a reasonable measure of distance, e.g., the trinocular distance \cite{HP10}.

\subsection{Baseline in polar direction}
\label{basetheta}

Consider the same model in Section \ref{SpheMassShell}. Let us choose the $z$ axis joining the origin at the shell center $C$ and the pointlike emitter $e$ so that $\theta_e=0$ in bookkeeper coordinates in (\ref{SphereMetric}), then specify the location of the observer $O$ as $(r_o, \theta_o, \varphi_o)$ (Figure \ref{plotDefdphi}). As shown in Appendix \ref{RevGeoEq}, each light ray connecting the emitter and the observer is always in an $r\theta$ plane of constant $\varphi$ by symmetry, and here, the constant is $\varphi_o$.
Suppose the baseline of the localized observer around $O$ is going in the $\theta$ direction about the $z$ axis in this $r\theta$ plane with a sufficiently small length such that the whole baseline can be considered as on the sphere of $r=r_o$. In Figure \ref{plotDefdphi} one can see that a light ray of the impact parameter $b$ emitted by a point source $e$ and received by the observer $O$ at the angle of arrival $\theta_a$ has the tangent line ${\bf r}(s) = [r_o \cos \theta_o + s \cos(\theta_o - \theta_a+\pi), r_o \sin \theta_o + s \sin (\theta_o - \theta_a+\pi)]$ at $O$ ($s\in {\bf R}$ and ${\bf r}(0)=(r_o, \theta_o)$) in the rectangular coordinates $[z, x]=[r \cos \theta, r \sin \theta]$ on the $r\theta$ plane of $\varphi=\varphi_o$. 
Suppose another light ray from the same emitter $e$ but with a slightly different value of the impact parameter $b_\varepsilon$, where $\varepsilon$ is a small parameter and $b_\varepsilon\to b$ as $\varepsilon\to 0$, is lying on the same $r\theta$ plane of $\varphi=\varphi_o$ and reaches a slightly different point $(r_o, \theta_o^\varepsilon)$ on the baseline from $O$. The tangent line of the second light ray at $(r_o, \theta_o^\varepsilon)$ is ${\bf r}_\varepsilon(s_\varepsilon) = [r_o \cos \theta_o^\varepsilon + s_\varepsilon\cos(\theta_o^\varepsilon - \theta_a^\varepsilon+\pi), r_o \sin \theta_o^\varepsilon + s_\varepsilon \sin (\theta_o^\varepsilon - \theta_a^\varepsilon+\pi)]$, $s_\varepsilon\in {\bf R}$. These two tangent lines intersect at
\begin{equation}
  s = \frac{r_o\left( \sin\left[\theta_o^\varepsilon -(\theta_o^\varepsilon - \theta_a^\varepsilon)\right]
	    -\sin\left[\theta_o -(\theta_o^\varepsilon - \theta_a^\varepsilon)\right]\right)}
			{\sin\left[(\theta_o^\varepsilon - \theta_a^\varepsilon)-(\theta_o - \theta_a) \right]} = 
			\frac{r_o \theta'_o \cos\theta_a}{\theta'_o - \theta'_a} + O(\varepsilon),  \label{sfordtheta}
\end{equation}
where $\theta'_o \equiv \lim_{\varepsilon\to 0} \left( \theta_o^\varepsilon - \theta_o \right)/\varepsilon$ and $\theta'_a \equiv \lim_{\varepsilon\to 0} \left( \theta_a^\varepsilon - \theta_a \right)/\varepsilon$.
Thus, the binocular distance determined by the infinitesimal baseline in the $\theta$ direction for the observer $O$ would be
\begin{equation}
  d_\parallel \equiv N(\theta_a)r_o \cos\theta_a\frac{\theta'_o}{\theta'_o - \theta'_a}, \label{dprltheta}
\end{equation}
where a normalization factor $N(\theta_a) \equiv \sqrt{A_o^{-1}\cos^2\theta_a +\sin^2\theta_a}$ is introduced to match the radar distance $\tilde{r}$ around the observer $O$ (cf. Section \ref{lumin}). 
Given $\theta_a$ in (\ref{thetaab}), and notice that $\theta_a^\varepsilon$ has the same form as $\theta_a$ except $b$ is replaced by
$b_\varepsilon = b + \varepsilon b' + O(\varepsilon^2)$, one has
\begin{equation}
  \theta'_a = \frac{b' \sqrt{A_o} \, r_o^2}{\sqrt{r_o^2 -b^2}\left[ A_o (r_o^2 - b^2)+ b^2 \right]}. \label{thetaaprime}
\end{equation}

\begin{figure}
\includegraphics[width=5.03cm]{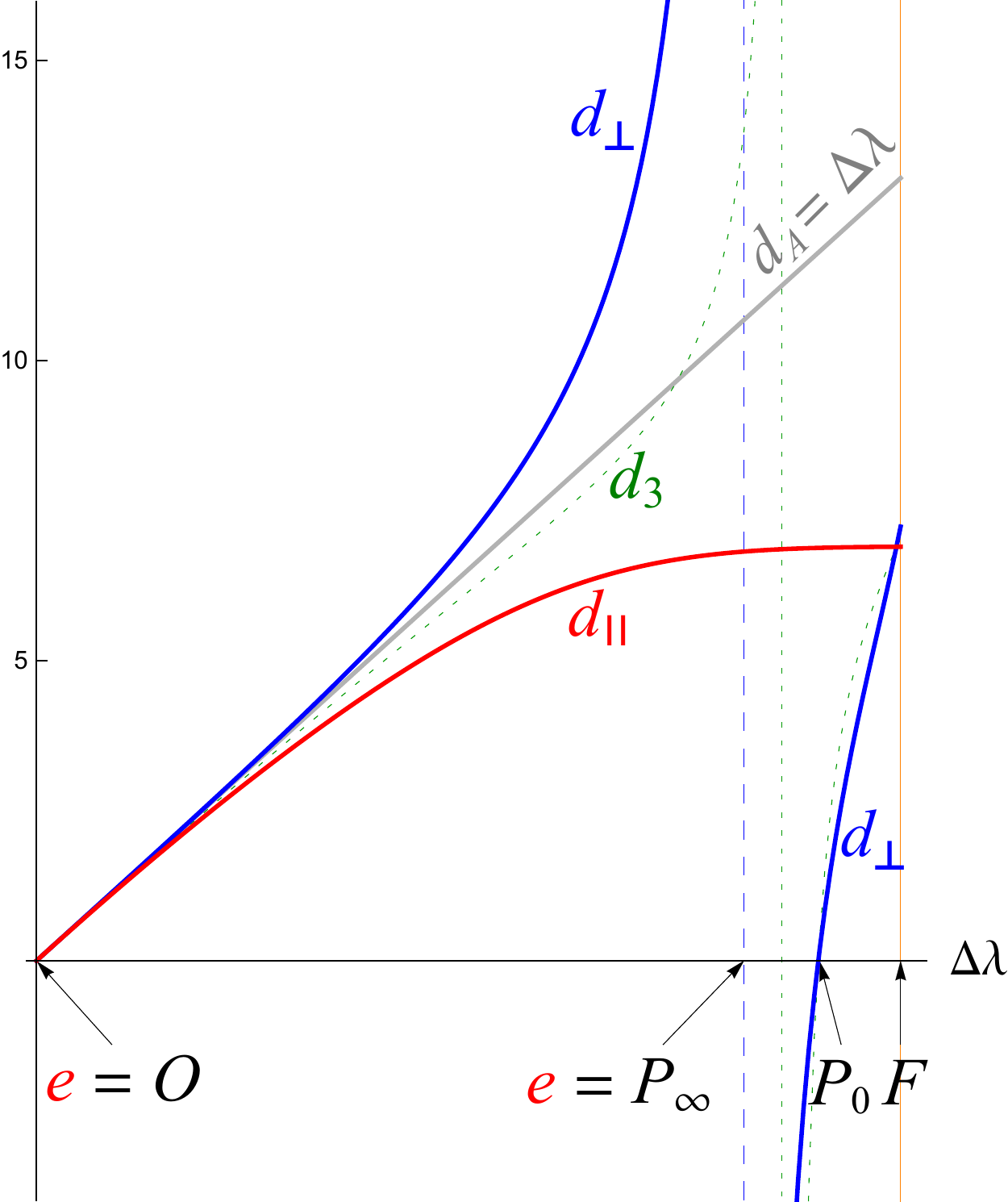}
\includegraphics[width=5cm]{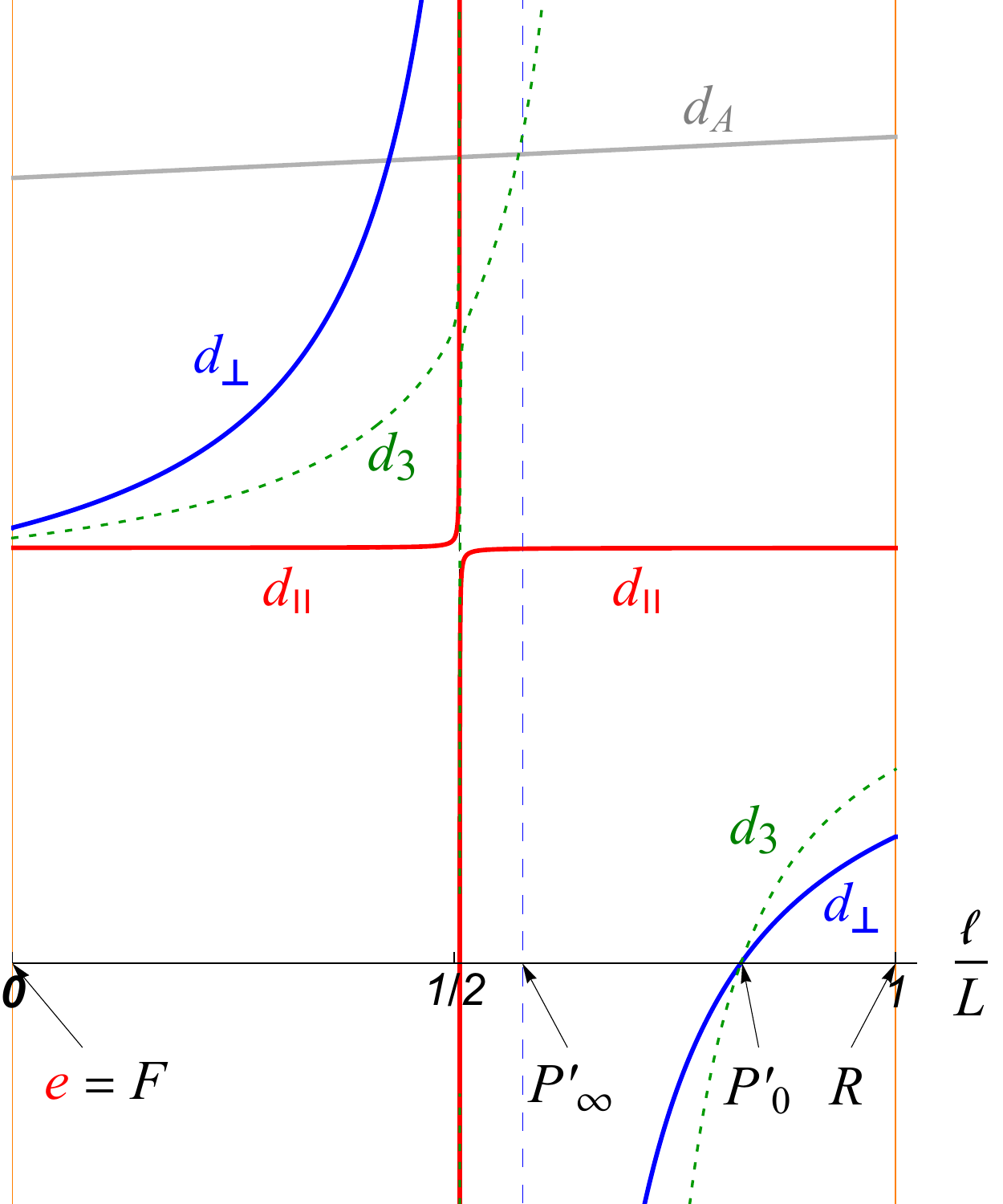}
\caption{Comparison of the affine distance $d_A$ (gray) with the binocular distances $d_\parallel$ (red), $d_\perp$ (blue), and the trinocular distance $d_3$ (green dotted) of the emitter $e$ for the observer $O$ as $e$ is brought from $O$ along the light ray (black) in Figure \ref{plotDefdphi} for $e$ outside (left plot) and inside the shell (right). Here, $\ell = \sqrt{A_o/A_s}\,\Delta\lambda$.}
\label{DisPrlx}
\end{figure}

Consider a light ray started at $R$ and received by the far observer $O$ represented as the black path in Figure \ref{plotDefdphi}, where $R$ is on the rear surface of the shell with respect to the affine distance for the observer. If we put a point source $e$ at another point than $R$ on the same black path, then one of the light rays emitted by $e$ can go along the black path to reach the observer $O$. Let us put an emitter $e$ at $O$, and then slowly bring $e$ away from $O$ toward $R$ along the black path in Figure \ref{plotDefdphi}, while the $z$ axis from $C$ pointing to $e$ varies following the move of $e$. When the emitter $e$ is still outside the shell, 
given $\theta_o$ in (\ref{thetaofr}), one has
\begin{equation}
  \theta'_o = b' \partial_b  \int_{r^{}_e}^{r_o} \frac{-b\,d r}{\sqrt{A_o r^4-b^2 r^2+2Mb^2 r}}
	= -\int_{r^{}_e}^{r_o}\frac{b' A_o r^4 dr}{\left( A_o r^4 - b^2 r^2 +2M b^2 r\right)^{3/2}}, 
	\label{thetaoprimeOut}
\end{equation}
since ($r_e$, $\theta_e$) are fixed. 
Substitute (\ref{thetaoprimeOut}) and (\ref{thetaaprime}) into (\ref{dprltheta}), one finds that the factor $b'$ cancels. 
As the emitter is brought away from $O$ along the black path in Figure \ref{plotDefdphi},
the binocular distance $d_\parallel$ is a monotonically increasing function of the affine distance $d_A=\Delta\lambda$ for the observer, as shown in Figure \ref{DisPrlx} (left). When $d_A$ is small, we have $d_\parallel \approx d_A$. As the path of the light ray from $e$ to $O$ starts to bend on the $r\theta$ plane, the value of $d_\parallel$ becomes less than $d_A$. As the emitter $e$ goes toward the spherical shell further, $d_\parallel$ tends to saturate but still increases all the way to the point that $e$ touches $F$ on the front surface of the shell with respect to $d_A$ for the observer.  
Indeed, as the emitter $e$ is brought away from the observer $O$ along the black path in Figure \ref{plotDefdphi}, $\theta'_a/b'$ is a positive constant from (\ref{thetaaprime}), and the integrand of $(-\theta'_o/b')$ is positive definite from (\ref{thetaoprimeOut}), so $(-\theta'_o/b')$ is increasing as $r_o-r_e$ increases from zero. Rewrite (\ref{dprltheta}) into $d_\parallel = N(\theta_a) r_o \cos\theta_a \times (-\theta'_o/b')/[(-\theta'_o/b') + (\theta'_a/b')]$, and the behavior of $d_\parallel$ described above is obvious.

When the emitter $e$ gets into the spherical shell, from (\ref{thetaofr0}), one has
\begin{eqnarray}
  \theta_o &=& \theta_e - \int_F^e \frac{b\sqrt{A_s/A_o}\, dr}{\sqrt{r^4 - r^2 b^2 (A_s/A_o)}} -  
	    \int_{O}^{F} \frac{b\,d r}{\sqrt{A_o r^4-b^2 r^2+2Mb^2 r}} \nonumber\\
  &=& \theta_e + \theta_d(b) -\theta_{\rm in}(b) -\int_{r_s}^{r_o} \frac{b\,d r}{\sqrt{A_o r^4-b^2 r^2+2Mb^2 r}} \label{thetaoIn}         
\end{eqnarray}
where $\theta_e=0$ in our choice of the $z$ axis, $\theta_{\rm in}=\sin^{-1}(-b\sqrt{A_s/A_o}/r_s)$ 
is the angle of incidence given in (\ref{thIthT}) ($|\theta_{\rm in}|=\angle CFe=\angle CRe$ in Figure \ref{plotDefdphi}),
and $\theta_d$ is the angle of departure (\ref{defthetad}) ($|\theta_d| = \pi - \angle CeF = \angle CeR$ in Figure \ref{plotDefdphi}). 
Note that $\theta_d =\theta_{\rm in}$ when $e=F$ and $\theta_d =-{\rm sgn}(b)\pi-\theta_{\rm in}$ when $e=R$ with ${\rm sgn}(b)=1$ for $b\ge 0$ and $-1$ for $b<0$. Then,
\begin{equation}
  \theta'_o = b'\left\{ \kappa(\ell) -\int_{r_s}^{r_o}\frac{A_o r^4 dr}{\left( A_o r^4 - b^2 r^2 +2M b^2 r\right)^{3/2}}\right\}
	\label{thetaoprimeIn}
\end{equation}
after some algebra, where 
\begin{equation}
  \kappa(\ell) \equiv \frac{2}{L}\sqrt{\frac{A_s}{A_o}}\, \frac{\ell}{\ell-\frac{L}{2}}  \label{kappaell}
\end{equation}
with the depths $\ell(b)$ and $L(b)$ defined in (\ref{defellL}).

Inserting (\ref{thetaaprime}) and (\ref{thetaoprimeIn}) into (\ref{dprltheta}), $d_\parallel$ reads 
\begin{equation}
     d_\parallel = N(\theta_a)r_o\cos\theta_a \frac{\kappa(\ell) +(\theta'_o/b')|_F}
		{\kappa(\ell)+ (\theta'_o/b')|_F - (\theta'_a/b')}, \label{dprlthetaIn}
\end{equation}
where $\left.(\theta'_o/b')\right|_F$ denotes the value of $\theta'_o/b'$ when $e$ is at $F$, 
and still, $\theta'_a/b'$ is a positive constant independent of $\ell$ or $\lambda$ from (\ref{thetaaprime}).

When $e$ is started at $F$, $\kappa(\ell=0)=0$, $\left.(\theta'_o/b')\right|_F <0$ from (\ref{thetaoprimeIn}), and so both the denominator and the numerator of (\ref{dprlthetaIn}) are negative. Moving the emitter $e$ from $F$ toward $R$ along the black path of light ray in Figure \ref{plotDefdphi}, $\ell$ increases, and $\kappa$ runs from $0$ down to $-\infty$ as $\ell \to \frac{L}{2}-$. When $e$ is passing through the middle point between $F$ and $R$, the observer would perceive $d_\parallel=N(\theta_a)r_o\cos\theta_a$. Then, as $e$ is brought from the middle point to $R$, $\kappa$ runs from $+\infty$ down to $(4/L)\sqrt{A_s/A_o}$. If $\left|(\theta'_o/b')\right|_F >(4/L)\sqrt{A_s/A_o}$, which is always true when $r_o$ is sufficiently large and the shell radius $r_s$ is close enough to $2M$ so that $A_s/A_o$ is sufficiently small, on the way that $\kappa$ drops from positive infinity down, $\kappa$ first meets the value of $(\theta'_a/b')-\left.(\theta'_o/b')\right|_F$, causing a divergence of $d_\parallel$ at that point. Then, $d_\parallel$ increases from negative infinity up as $\ell$ keeps increasing, until $\kappa$ passes through the value of $-\left.(\theta'_o/b')\right|_F$ where $d_\parallel =0$. After that point, $d_\parallel$ becomes positive and still increases, all the way to the point that $e$ arrives at $R$ where $\ell=L$, $\kappa = (4/L)\sqrt{A_s/A_o}$, $\left.(\theta'_a/b')\right|_R > 0$, and remarkably $\left. d_\parallel \right|_R < \left. d_\parallel \right|_F$.

In Figure \ref{DisPrlx} (right), one can see that, when $r_s$ is sufficiently close to $2M$, the slope of $d_\parallel(\ell)$ is approximately zero in almost the whole domain $0 \le \ell \le L$ inside the shell except the neighborhood of the singularity around the middle point $\ell=L/2$. This is because $\kappa(\ell)$ in (\ref{kappaell}) is suppressed by the factor $\sqrt{A_s/A_o}\to 0$ 
when $\ell$ is not very close to $L/2$. Thus, it would become more and more difficult for the observer to resolve the depth $\ell$ of the emitter $e$ from $d_\parallel$ as $r_s\to 2M$, unless $e$ happens to be located around the singularity of $d_\parallel(\ell)$. 

When the interior emitter $e$, the shell center $C$, and the observer $O$ are lying on the same straight line, i.e., the $z$ axis on the $r\theta$ plane, one has $\theta_o =0$ or $\pi$, and the direct image of the emitter $e$ concentrated around $\theta_a = 0$ for the observer  is produced by the light rays of $b\approx 0$. In particular, the binocular distance of this direct image for the observer $O$ with an infinitesimal baseline in the polar direction is determined by the light rays of infinitesimal but non-vanishing $b$. Taking the limit $b\to 0$, Eq.(\ref{dprlthetaIn}) can be expressed in closed form,
\begin{equation}
 d_\parallel = \frac{1}{\sqrt{A_o}} \left[r_o - \frac{r_s(r_s-\ell)}{r_s-(1-\sqrt{A_s})\ell}\right] \label{dthb0}
\end{equation}
with $0 < \ell < 2r_s$. The behavior of the above $d_\parallel(\ell)$ is similar to the one in Figure \ref{DisPrlx} (right).
For small $\ell$, the value of $d_\parallel(\ell)$ in (\ref{dthb0}) is close to the affine distance $d_A(\ell) = (r_o-r_s+\sqrt{A_s}~\ell)/\sqrt{A_o}$ in (\ref{affdistIn}) with the light ray of $b=0$.

For a near observer inside the photon sphere, the light rays observed at the angles of arrival $|\theta_a| < \theta_a^c$ behave like those in the case with the far observer outside the photon sphere. The light rays coming at the angles of arrival $\theta_a \in (\theta_a^c, \pi - \theta_a^c)$, on the other hand, are trapped in the photon sphere and may have been going out of and into the spherical shell several times from the emitter inside the shell. Since $\theta(\lambda)$ is monotonic in affine parameter $\lambda$, compare (\ref{thetaoIn}) and (\ref{thetaoprimeIn}), one can see that
\begin{equation}
  \theta'_o = b'\left\{ \kappa(\ell) + n\left[ \kappa(L) -
	2\int_{r_s}^{r_{\rm max}} \frac{A_o r^4 dr}{\left( A_o r^4 - b^2 r^2 +2M b^2 r\right)^{3/2}}\right]
	-\int_{F}^{O}\frac{A_o r^4 dr}{\left( A_o r^4 - b^2 r^2 +2M b^2 r\right)^{3/2}}\right\},
	\label{thetaoprimeIn3M}
\end{equation}
for the light ray from $e$ to $O$ going out of and then into the shell for $n$ times before the last section to the observer $O$ totally outside the shell. Here, $\int_F^O = \int_{r_s}^{r_o}$ for $|\theta_a|\le\pi/2$, and $\int_F^O = \int_{r_s}^{r_{\rm max}} + \int_{r_o}^{r_{\rm max}}$ for $|\theta_a| > \pi/2$. Inserting the above $\theta'_o$ into (\ref{dprltheta}), the perceived images of the emitters along this kind of light rays in terms of the binocular distance $d_\parallel$ would distribute in a multi-layer structure similar to those in terms of the affine distance $d_A$ (e.g., Figure \ref{DisAinPS}). If we move an emitter $e$ from $O$ along such a light ray, since the contribution by the integrals in (\ref{thetaoprimeIn3M}) is negative definite, $d_\parallel$ will increase as $d_A$ increases when $e$ is outside the spherical shell, and will behave like the one in Figure \ref{DisPrlx} (right) when $e$ is passing through the inside of the shell. 
As $r_s \to 2M$, all the $\kappa$ functions are suppressed except the singularities around the shell center, and so the perceived thickness of each image layer of the interior emitters goes to zero in terms of $d_\parallel$, too. The distance $d_\parallel$ of the $n$-th image layer  increases monotonically as $n$ increases, and then saturates to the value $N(\theta_a)r_o\cos\theta_a$ as $n\to\infty$.   
If the observer is not too close to the shell surface, the gaps between the image layers would always be clear, and the direct images at $|\theta_a|\le \theta_1$ in the nearest image layer would be able to be identified easily by the near observer in terms of $d_\parallel$, too.

The values of $\left. -(\theta'_o/b')\right|_F$ of the light rays observed around the boundary of the whole star image [see the statement above (\ref{largestangle})], $\tilde{\theta}_a = \tilde{\theta}_a^c$ for a far observer and $\pi-\tilde{\theta}_a^c$ for a near observer, diverge to positive infinity due to the singularity of the integrand at $r=3M$ in (\ref{thetaoprimeIn}), while $\left.(\theta'_a/b')\right|_F$, $\left.\kappa\right|_F$, and $\left.\kappa\right|_R$ of the same light rays are finite. 
So, the binocular distances $d_\parallel$ in (\ref{dprlthetaIn}) for the front and rear surfaces converge to the same value $N(\theta_a^c)r_o \cos\theta_a^c$ at the boundary of the star image [Figure \ref{emitterarray} (lower-right)]. 
For all $\tilde{\theta}_a < \tilde{\theta}_a^c$, $\left. d_\parallel \right|_R$ is less than $\left. d_\parallel \right|_F$, so the front surface of the shell with respect to the affine distance would be positioned behind the rear surface in terms of $d_\parallel$ for the observer. The images of the interior emitters between the front surface and roughly the shell center in bookkeeper coordinates would be located behind the perceived front surface, as indicated in Figure \ref{DisPrlx} (right), while the apparent locations of the interior emitters at even deeper $\ell$ would be somewhere in front of the perceived rear surface, and so the star looks inside out. 
Most of those emitter images look very close to the shell surfaces, as shown in Figure \ref{emitterarray} (lower-right).
As the shell radius $r_s$ goes to the Schwarzschild radius $2M$, the images of the shell surfaces and almost all the interior emitters squeeze to a membrane in terms of $d_\parallel$, except the emitters around the shell center $C$.

The zero value of the distance $d_\parallel$ for an emitter around the shell center does not imply that the observer would see the image of that emitter in the vicinity of the observer. In fact, the binocular distance $d_\parallel$ at a fixed angle of arrival $\theta_a=\bar{\theta}_a$ is ill defined around the zero of (\ref{sfordtheta}) or (\ref{dprlthetaIn}) where $\ell =\ell^{}_0 \approx\kappa^{-1}\left(-\left.(\theta'_o/b')\right|_F\right)$ from (\ref{dprlthetaIn}). The observer at ($r_o, \theta_o$) is a focal point of the light rays around the angle of arrival $\bar{\theta}_a$ from the interior emitter at depth $\ell_0$, and the vanishing value of $d_\parallel(\ell_0)$ is associated with a zero length of baseline, which is beyond our assumption of infinitesimal baseline. Suppose we introduce a small but finite baseline in the polar direction with two telescopes situated at its two ends and the observer $O$ sitting in the middle. Then, as the depth $\ell$ of an emitter is approaching $\ell_0$ along the interior section of the null geodesic, one of the telescopes would start to miss all the light rays in the neighborhood of the angle of arrival $\bar{\theta}_a$, and the binocular distance $d_\parallel$ could not be determined in the conventional binocular way, until $\ell$ gets to be sufficiently away from $\ell_0$ and the missing image re-appears. According to such a kind of image missing in one of the telescopes, anyway, the observer could still know that the location of the emitter is around the shell center $\ell \approx r_s$ rather than at other depths even when $r_s$ is very close to $2M$.

\subsection{Baseline perpendicular to polar direction}
\label{basephi}

With the same coordinate system chosen in Section \ref{basetheta}, suppose now the baseline of the localized observer at $(r_o,\theta_o, \varphi_o)$ is perpendicular to the plane of the light ray $\varphi=\varphi_o$ with a small length $r_o \delta\varphi$ extended in each direction on the sphere of $r=r_o$ along the great circle. Let the spatial part of bookkeeper coordinates of the two ends of the baseline be $O_\pm=(r_o, \theta_o +\Delta\theta, \varphi_o\pm \Delta\varphi)$. Then, $\Delta\varphi (\sim \delta\varphi)$ and $\Delta\theta(\sim (\delta \varphi)^2$ or less) are small for almost all $\theta_o$ except $|\theta_o-n\pi| \alt 2\delta\varphi$, $n\in {\bf Z}$. 
Given a light ray connecting the emitter $e$ and the observer $O$ in the $\varphi_o$ plane such as the black path in Figure \ref{plotDefdphi}, a rotation $\Delta\varphi$ or $-\Delta\varphi$ about the $z$ axis joining the shell center $C$ and the emitter $e$ will give another light ray from $e$ to ${\cal O}_\pm=(r_o,\theta_o,\varphi_o\pm\Delta\varphi)$, whose distance to the closer end of the baseline $O_\pm$ in bookkeeper coordinates is proportional to $\Delta\theta$. Thus, when we consider an infinitesimal baseline with $\delta\varphi\to 0$, for almost all $\theta_o$ except $|\theta_o-n\pi| \to 0$, we have $\Delta\theta \ll \Delta\varphi\to 0$ and so ${\cal O}_\pm \to O_\pm$. 
In Figure \ref{plotDefdphi} one can see that the tangent lines of the light rays at ${\cal O}_\pm$, which can be written as ${\bf r}(s)=[r_o \cos (-\theta_e+\theta_o) + s \cos(-\theta_e + \theta_o - \theta_a+\pi), r_o \sin (-\theta_e +\theta_o) + s \sin (-\theta_e+\theta_o - \theta_a+\pi)]$ ($s\in {\bf R}$, $\theta_e=0$) in terms of the rectangular coordinates on the $r\theta$ plane of each light ray ($\varphi = \varphi_o\pm \Delta\varphi$), will intersect at $X$ on the axis of rotation. Therefore, $s\, \sin \angle OXC = r_o \sin \angle OCX$, or
\begin{equation}
  s\, \sin [\pi-(\theta_e - \theta_o-\pi) -\theta_a] = r_o \sin (\theta_e-\theta_o-\pi)   
\end{equation}
equals the distance between $O$ and the $z$ axis in Figure \ref{plotDefdphi}, and the binocular distance determined by the infinitesimal baseline perpendicular to the $\varphi_o$ plane for the localized observer $O$ would be 
\begin{equation}
  d_\perp \equiv N(\theta_a)\frac{r_o \sin (\theta_e -\theta_o)}{\sin (\theta_e - \theta_o + \theta_a)}. \label{defdphi}
\end{equation}
Here, $\theta_a$ and $\theta_e -\theta_o$ on the $r\theta$ plane have been given in (\ref{thetaab}) and (\ref{thetaofr0}), respectively.

Our formula (\ref{defdphi}) does not work for $|\theta_o-\theta_e| = n\pi$, $n\in {\bf Z}$, when the emitter $e$, the shell center $C$, and the observer $O$ are lying on a straight line (the $z$ axis) in bookkeeper coordinates. In these cases, one has $\Delta\varphi = \pi/2$ and $|\Delta \theta | = \delta\varphi$, and the baseline perpendicular to the $\varphi_o$ plane is actually in the polar direction in the plane of $\varphi=\varphi_o\pm \pi/2$. Each telescope at the two ends of this baseline will see the single (for $\theta_a=0$) or double image (for $\theta_a\not=0$) of the emitter $e$ produced by the light rays going in the degenerate plane of $\varphi=\varphi_o\pm \pi/2$, and the binocular distance for each image measured by the observer using the above baseline would rather be the non-zero $d_\parallel$ determined on that plane, though (\ref{defdphi}) gives $d_\perp=0$ in these cases. 

As in Section \ref{basetheta}, let us put the emitter $e$ at the position of the observer $O$ and bring $e$ slowly away from $O$ along the black path in Figure \ref{plotDefdphi}, where $O$ is outside the photon sphere and the explicit expressions for $\theta_e -\theta_o$ in (\ref{thetaofr}) and (\ref{thetaoIn}) apply. One can see in Figure \ref{DisPrlx} (left) that initially the emitter would be perceived by the far observer $O$ as an image at a slowly growing binocular distance $d_\perp$ (blue curve), whose value is very close to the affine distance $d_A$ from $O$ to $e$. When the emitter $e$ is sufficiently close to $P_\infty$ in Figure \ref{plotDefdphi}, the value of $d_\perp$ starts to deviate from the affine distance $d_A$, and then diverges at $P_\infty$ where the $z$ axis joining the emitter $e$ and the shell center $C$ is parallel to the tangent line of the light ray at $O$, such that $\theta_e-\theta_o+\theta_a= n \pi$, $n\in {\bf Z}$, and the incoming light rays from the emitter $e$ are parallel to each other around the two ends of the observer's baseline.  
As the emitter $e$ is brought further from $P_\infty$ toward $F$, $d_\perp$ becomes negative and increases from negative infinity, 
while the focal point of the rays emitted from $e$ at different $\varphi$ is behind the observer and the tangents of the received rays will not intersect in front of the observer as the observer is facing to the center of the shell. 
When the emitter $e$ approaches $P_0$, which is exactly on the straight line joining the shell center $C$ and the observer $O$, 
$d_\perp$ goes to zero, and then (\ref{defdphi}) breaks down at $e=P_0$. 

As the emitter is moved farther away from $P_0$ toward $F$ in Figure \ref{plotDefdphi}, the distance $d_\perp$ would grow from zero to some positive value (Figure \ref{DisPrlx} (left)). Before the emitter $e$ arrives at $F$ on the front surface of the shell, 
the light rays outside the shell from $e$ to $O$ may contain zero (small $b$) to many ($b \to b_c$) intervals of negative $d_\perp$ similar to $(P_\infty, P_0)$, depending on the value of the impact parameter $b$ and thus the angle of arrival $\theta_a$.
 
Inside the shell in Figure \ref{plotDefdphi}, the black path crosses the straight line going through $C$ and parallel to $OX$ (gray dashed) at $P'_\infty$, and crosses the straight line joining $C$ and $O$ (black) at $P'_0$. The behavior of the binocular distance $d_\perp$ of the emitter $e$ around the interval $(P'_\infty, P'_0)$ is similar to those in $(P_\infty, P_0)$ for the observer $O$, as shown in Figure \ref{DisPrlx} (right). For every null geodesic of $b\not=0$ started at some $R$ on the rear surface, going across the interior of the star, 
and then arriving at the observer $O$ outside the shell, the section inside the shell must contain at least one of $P'_\infty$ and $P'_0$ if the shell radius $r_s$ is sufficiently close to the Schwarzschild radius $2M$ so that $r_{\rm min}\to 0$. 

\begin{figure}
\includegraphics[width=4cm]{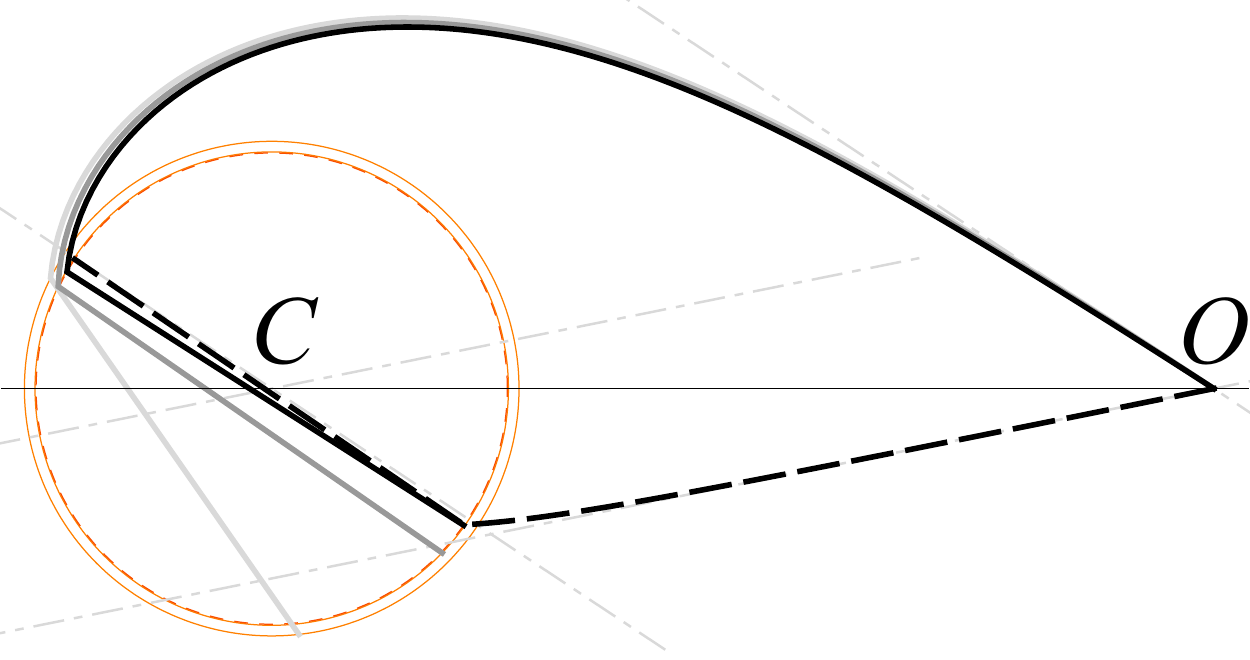}\hspace{.5cm}
\includegraphics[width=6cm]{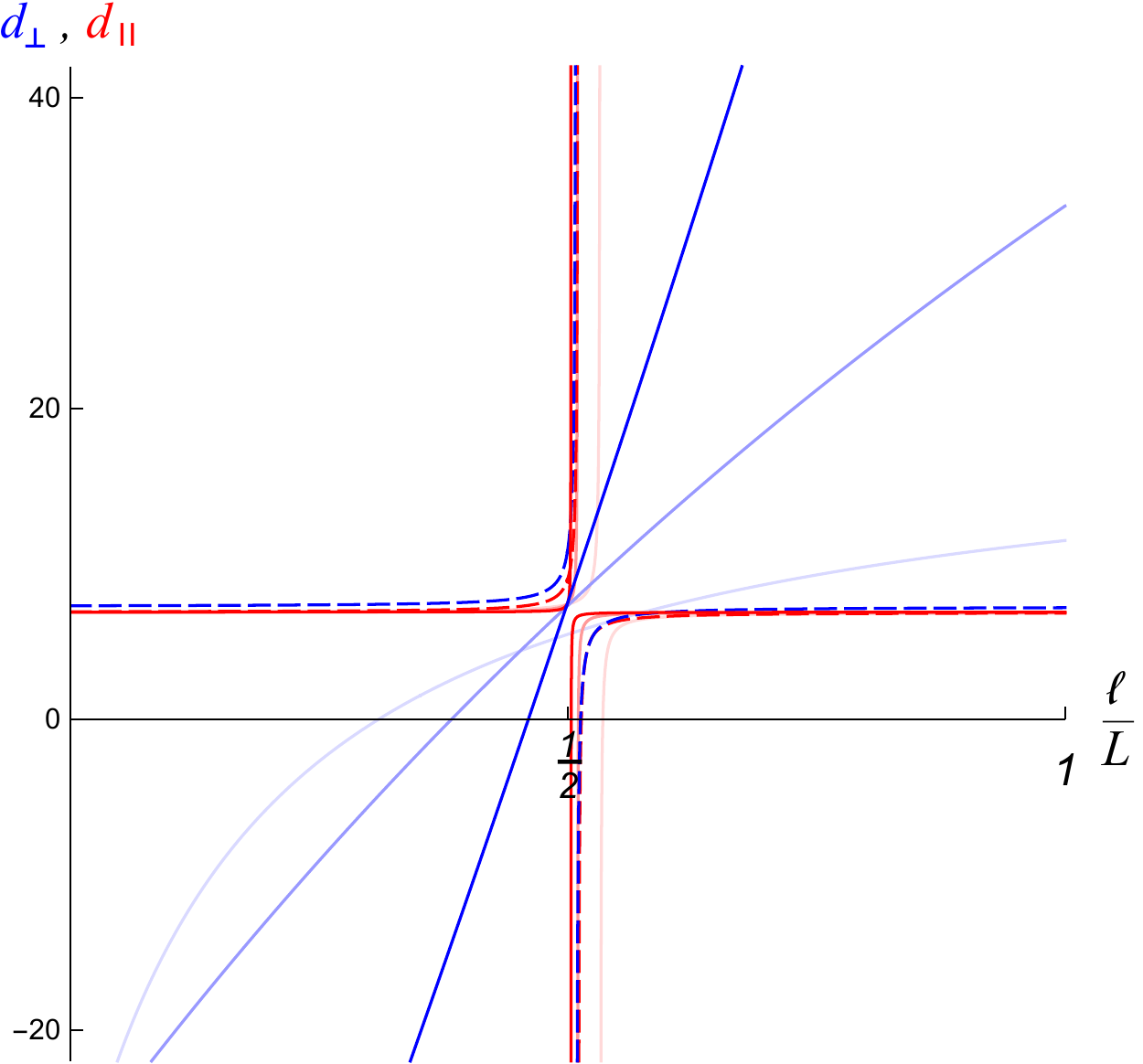}
\includegraphics[width=6cm]{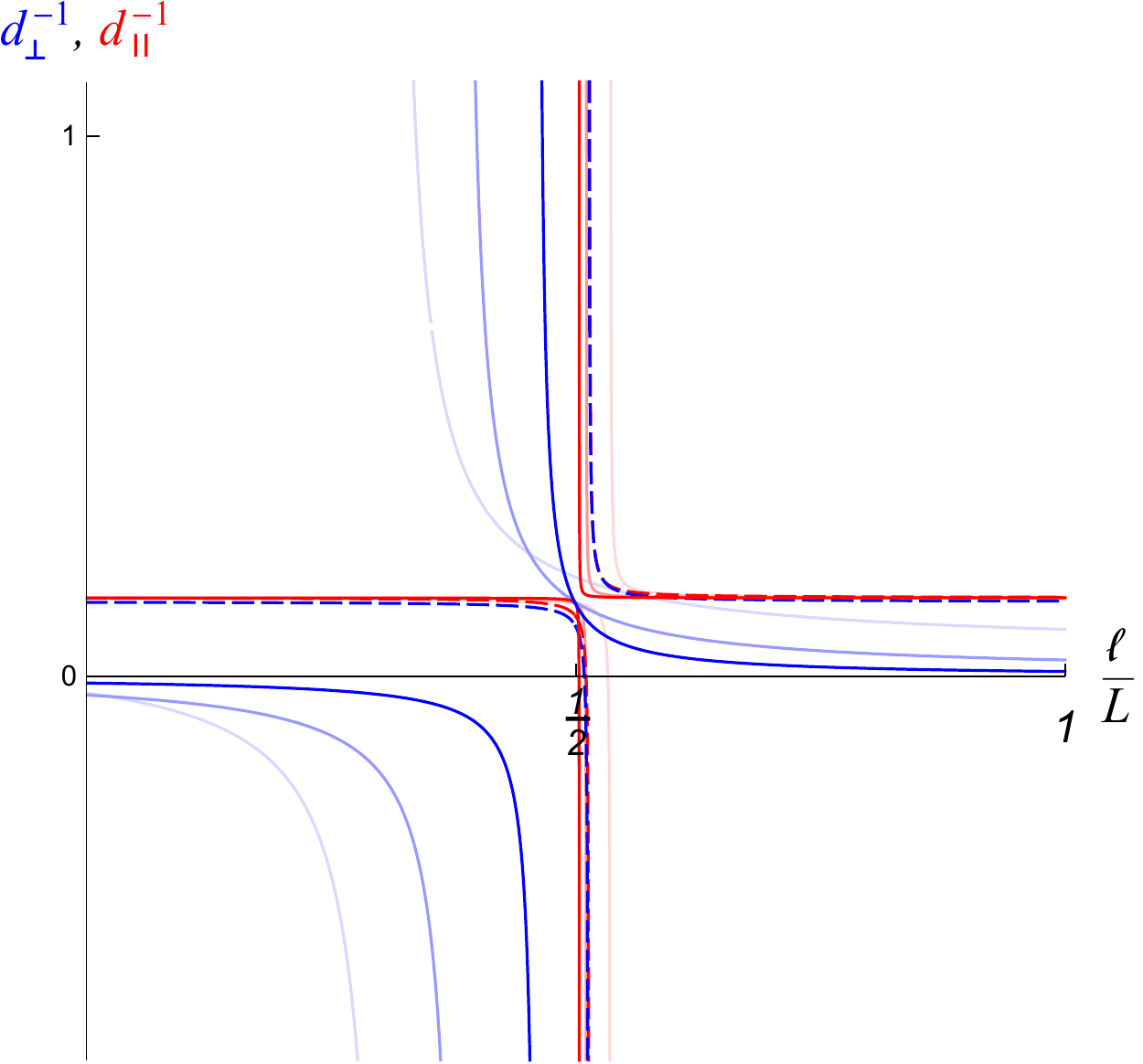}
\caption{Along the black, gray, or the very-light-gray solid curve in the left plot, $\theta_e-\theta_o+\theta_a \approx\pi$ for the emitter inside the spherical shell whose $d_\perp(\ell)$ (blue solid curves in the middle plot) for the observer becomes monotonically increasing with its depth $\ell$. This behavior is different from the normal cases (dashed curves, and the example in Figure \ref{DisPrlx}). Here, $r_s=2.1M$, $b=3.98548$ (very-light-colored curves), $r_s=2.01M$, $b=3.94182$ (light colored), $r_s=2.001M$, $b=3.898$ (black, blue and red), $r_s=2.001M$, $b=-1.37543$ (dashed curves in all plots, normal case), with other parameters the same as those in Figure \ref{DisPrlx}. In the middle plot, one can see that as $r_s\to 2M$ (from very-light blue to blue) the slope of $d_\perp(\ell)$ increases and $\ell$ appears to be more easily resolvable. However, in the right plot $d^{-1}_\perp$ goes to zero in most values of $\ell$ inside the shell in the same limit, and so the resolution of depth $\ell$ from the directly measurable quantity $\Delta\varphi_a\propto d^{-1}_\perp$ is actually suppressed. 
We put $d_\parallel$ and $d^{-1}_\parallel$ (red curves; see Section \ref{basetheta}) in each cases for comparison.}
\label{anomaly}
\end{figure}

\begin{figure}
\includegraphics[width=5.5cm]{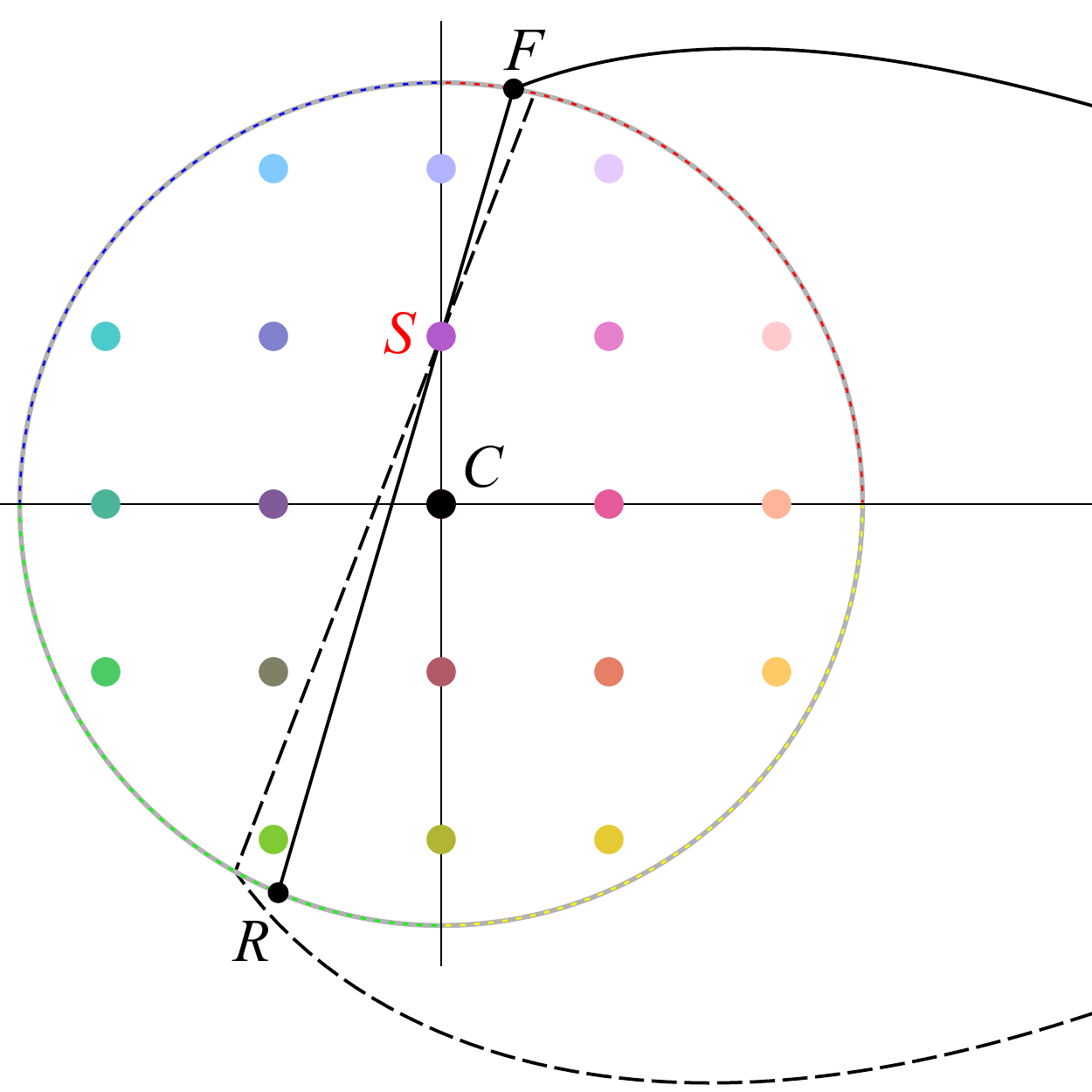}\hspace{1cm}
\includegraphics[width=5.5cm]{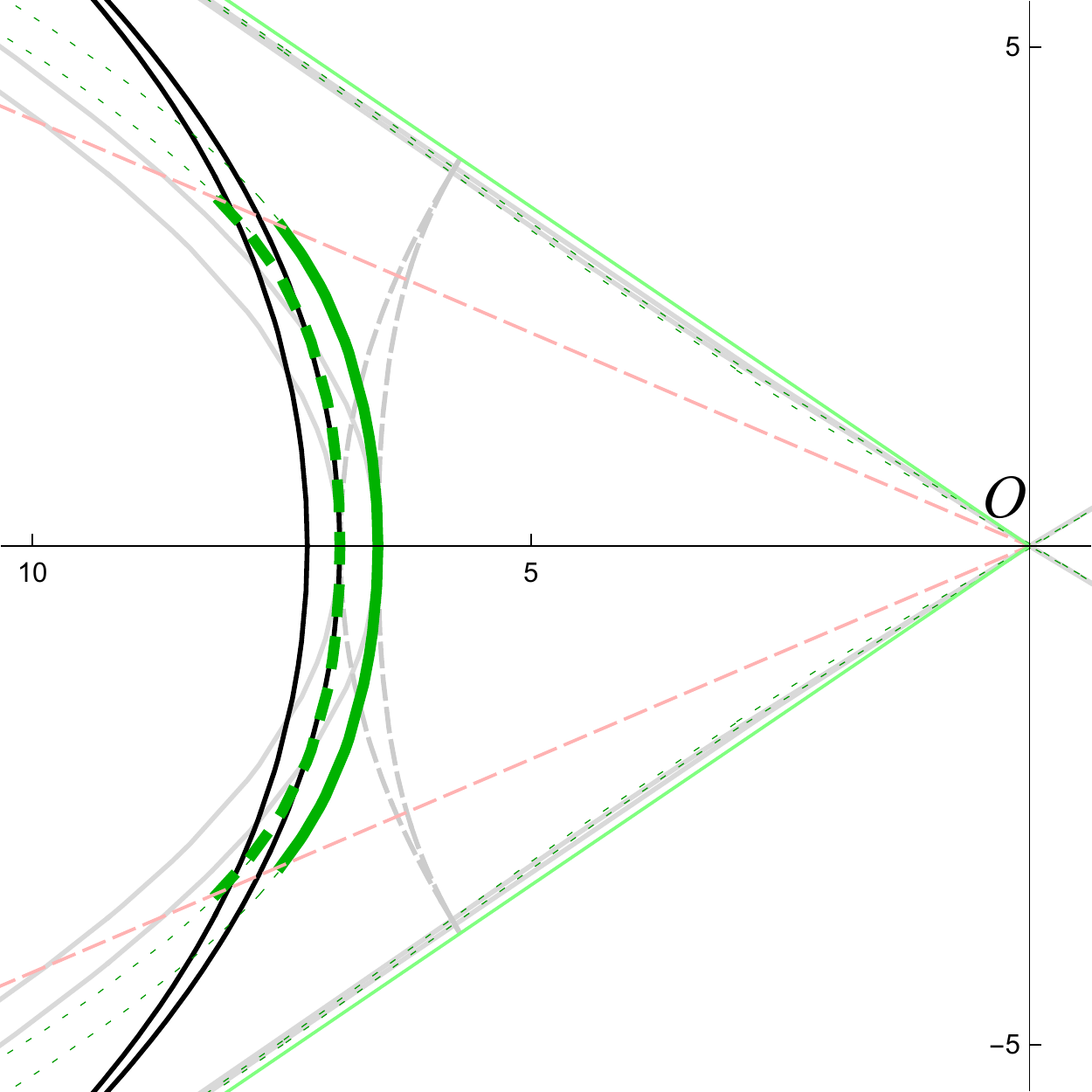}\\ \vspace{.5cm}
\includegraphics[width=5.5cm]{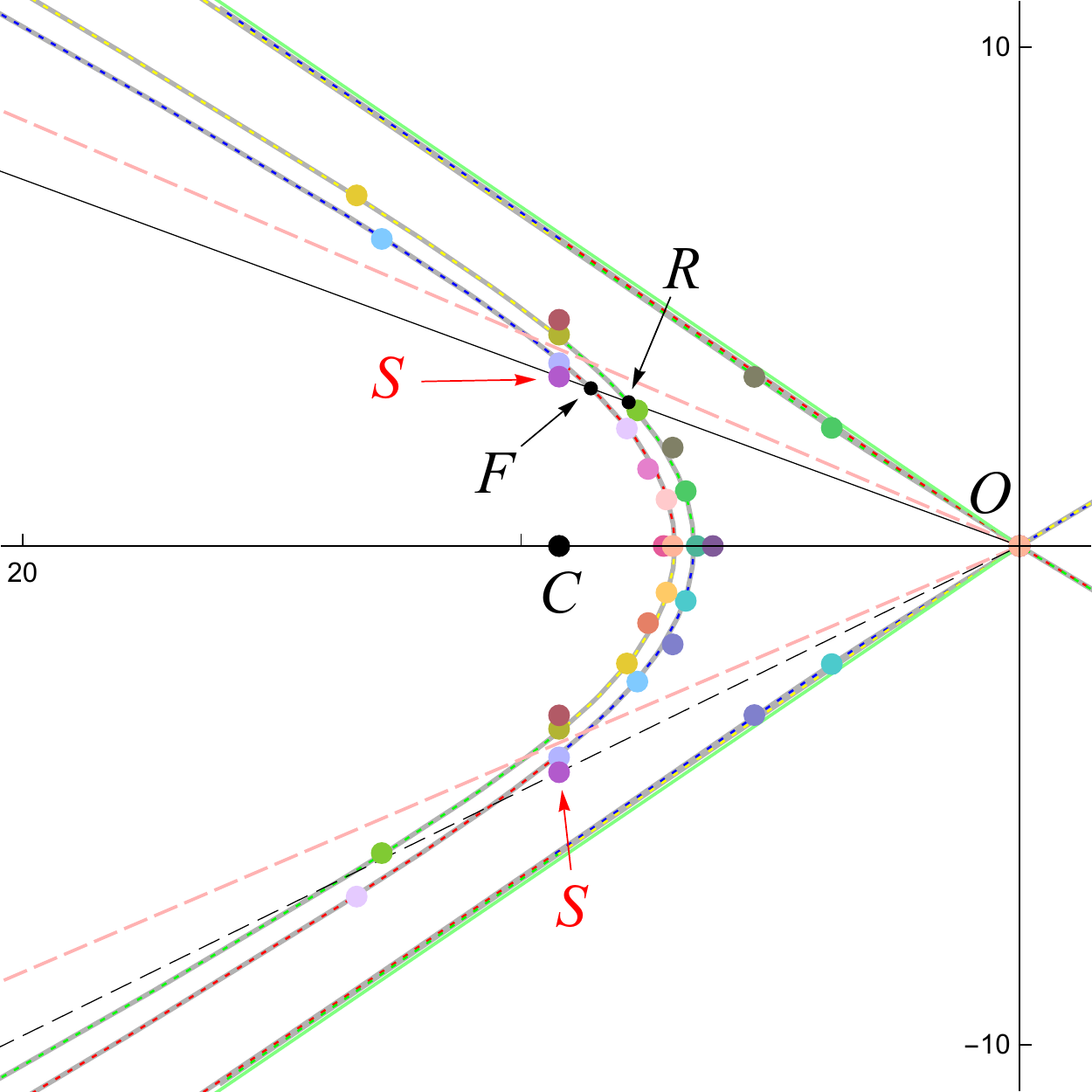}\hspace{1cm}
\includegraphics[width=5.5cm]{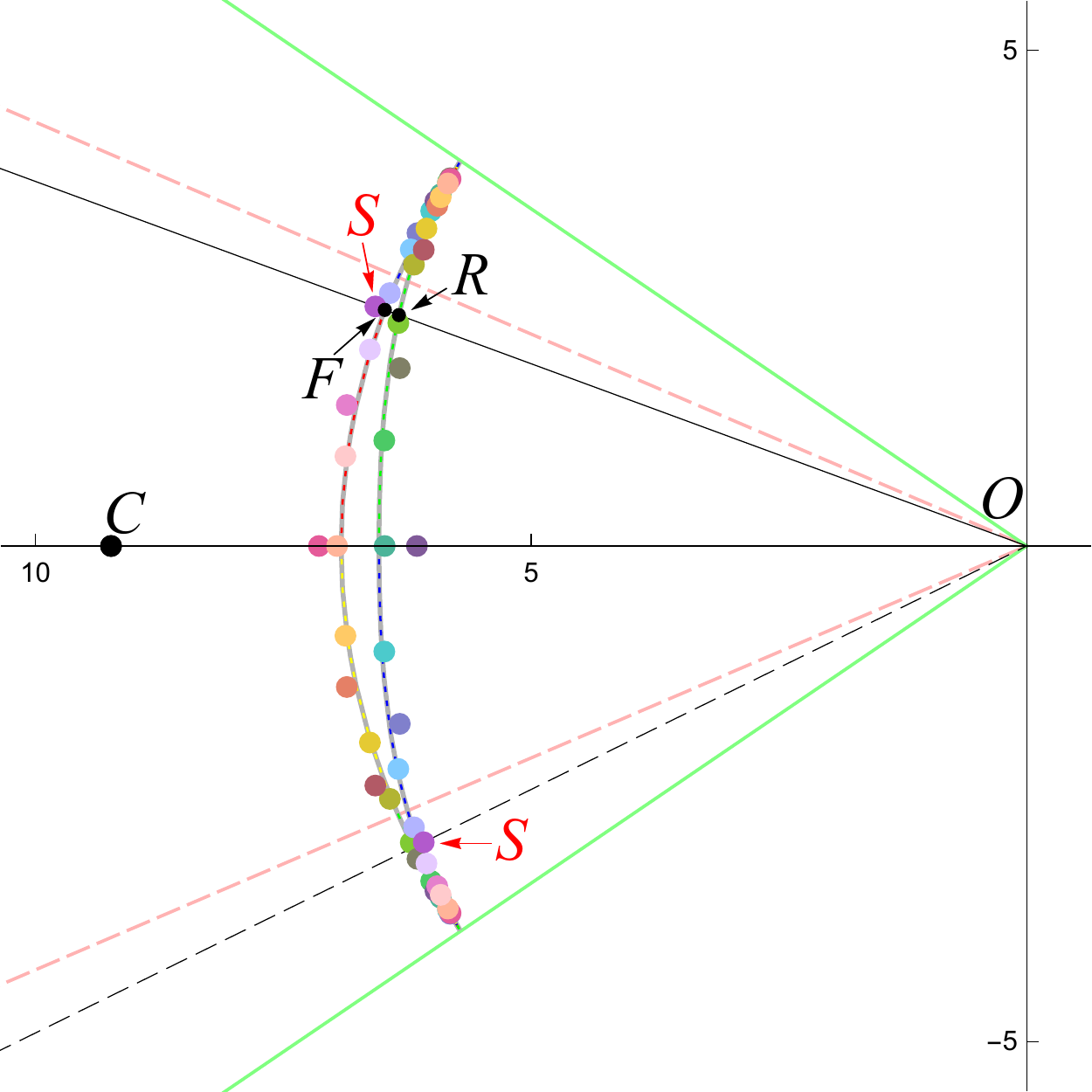}
\caption{ (Upper-left) An array of emitters on the $r\theta$ plane of some constant $\varphi$, distributed evenly inside the spherical shell (gray-based circle) of $r_s=2.01M$ and $M=1$. The far observer is localized at $r_o=8M$, 
the same as Figure \ref{theta1}. The colored spots represent the positions of emitters here and the corresponding images in the lower row, where the shapes or Doppler effect of the images is not considered. 
(Lower-left) Observed $(d_\perp, \tilde{\theta}_a)$ of the direct and first indirect images of the interior emitters on the $\tilde{r}\tilde{\theta}$ plane. The green and red-dashed straight lines mark the angles of arrival $\tilde{\theta}_a = \tilde{\theta}_c$ and $\tilde{\theta}_a = \tilde{\theta}_1$, respectively. The direct image of emitter $S$ (tracing along the black-solid ray with $\tilde{\theta}_a<\tilde{\theta}_1$ in the upper-left plot) and its first indirect image (along the black-dashed ray with $\tilde{\theta}_a > \tilde{\theta}_1$) are specified. (Lower-right) Observed $(d_\parallel, \tilde{\theta}_a)$ of the direct and first indirect images of the emitters on the $\tilde{r}\tilde{\theta}$ plane. (Upper-right) Comparison of the trinocular distance $d_3$ (green) and the affine distance $d^{}_A$ (black) of the shell surfaces for the observer at $O$. For small angles of arrival $\tilde{\theta}_a$ ($\alt \tilde{\theta}_1$ in this example), the maps of the front surface of the shell in terms of $d_A$ (black) and $d_3$ (thick green dashed) almost coincide, while the rear surface with respect to $d^{}_A$ appears in front of the ``front" surface when perceived in $d_3$ (thick green solid). The gray solid and dashed curves are the shell surfaces in $d_\perp$ and $d_\parallel$ which have been shown in the lower row.} 
\label{emitterarray}
\end{figure}

When the emitter $e$ is right at the point $F$ on the front surface, the observer $O$ would perceive $d_\perp|_F =N(\theta_a)\sin(\theta_F-\theta_o)/\sin(\theta_F-\theta_o+\theta_a)$. As we bring the emitter $e$ from $F$ to $R$ in Figure \ref{plotDefdphi},  
the depth $\ell$ increases from zero to $L$, and the behavior of the angle $\theta_e-\theta_o$ in (\ref{thetaoIn}) depends only on the angle of departure $\theta_d$ given in (\ref{thetad}) ($\theta_e=0$, and $\theta_{\rm in}$ is a constant of $\ell$).
When the emitter arrives at $R$ on the rear surface for the observer, the observer would see $d_\perp|_R = N(\theta_a) 
\sin(\theta_F-\theta_o + {\rm sgn}(b)\pi+2\theta_{\rm in})/\sin(\theta_F-\theta_o+ {\rm sgn}(b)\pi+2\theta_{\rm in}+\theta_a) 
= d_\perp|_F + 2 \theta_{\rm in} \times N(\theta_a)\sin\theta_a/\sin^2(\theta_F-\theta_o+\theta_a) +O(\theta_{\rm in}^2)$. 
In the cases with $\sin^2(\theta_F-\theta_o+\theta_a)$ not very close to zero, called the ``normal" cases in this paper [e.g., the blue dashed curve in Figure \ref{anomaly} (middle)), one has $d_\perp|_F > d_\perp|_R$ for sufficiently small $|\theta_{\rm in}|$ ($=\angle CFR$ 
in Figure \ref{plotDefdphi}] since $\theta_{\rm in}\sin\theta_a<0$ for all $b\not=0$ from (\ref{thetaio}) and (\ref{thetaab}).
This implies that in the normal cases the image of the front surface of the shell with respect to the affine distance $d^{}_A$, 
when perceived by the observer in $d_\perp$, would be positioned behind the image of the rear surface. Moreover, in Figure \ref{emitterarray} (lower-left), one can see that the interior emitters situated between $F$ and $P'_\infty$ will be perceived farther behind the front surface, and the images of the emitters between $P'_0$ and $R$ will be seen in front of the rear surface of the shell in $d_\perp$. Thus, the star looks inside out in terms of $d_\perp$ similar to the observations in terms of $d_\parallel$. 

As $r_s\to 2M$, $|\theta_{\rm in}|$ goes to zero from (\ref{thetaio}) so that the images of the rear and front surfaces in term of $d_\perp$ merge. In this limit, $\theta_d(\ell)$ in (\ref{thetad}) behaves like a step function, $\theta_d(\ell) \to 0$ for $\ell < L/2$ and $\theta_d(\ell) \to -{\rm sgn}(b)\pi$ for $\ell > L/2$, and so $d_\perp$ is approximately a constant for almost all $\ell\in [0,L]$.  
This implies that in the normal cases, as the shell is about to form a black hole, the images of most interior emitters would squeeze around the images of the shell surfaces when perceived in $d_\perp$, and from these images, the observer $O$ would not be able to resolve the depth $\ell$ of the interior emitters. The only exceptions are those emitters in the core region around the shell center $C$, in which $P'_\infty$ and/or $P'_0$ as well as the sudden change of $\theta_e-\theta_o$ around $\ell=L/2$ occur. Those images may have the distances $d_\perp$ ranging from $-\infty$ to $\infty$ when perceived by the observer [Figure \ref{emitterarray} (lower-left)].

Anomalous behavior occurs in the cases with $\theta_F-\theta_o+\theta_a\approx n\pi$, $n\in {\bf Z}$, when a small change in $\theta_e-\theta_o$ with $\ell$ would be amplified by the denominator in (\ref{defdphi}), and so the observer appears to be able to resolve $d_\perp(\ell)$. In Figure \ref{anomaly}, we show an example of such light rays. In the left plot, one can see that the sections of such light rays inside the shell are roughly parallel to the tangent lines of the same light rays at the observer $O$. Figure \ref{anomaly} (middle) shows that the distances $d_\perp$ of the emitters located along the sections of those light rays inside the shell are monotonically increasing with the depth $\ell$ (blue solid curves) and very different from the normal behavior mentioned above (blue dashed curve). As $r_s\to 2M$, the slope $\partial_\ell d_\perp$ increases, and the depths of the emitters along these rays appear to be resolvable even more easily around each window of $\theta_a(b)$ where $\theta_F-\theta_o+\theta_a\approx n\pi$ for the observer. 
One may be tempted to conclude that the observer can use these windows and change the location $(\theta_o, \varphi_o)$ to scan the whole interior of the shell, or compare with the data from other observers to get the depth information inside the shell. Nevertheless, these ideas would not be practical because a localized observer does not directly measure $d_\perp$. 
Rather, the observer infers $d_\perp = W/\Delta\varphi_a$ with the length of the baseline $W$ after directly measuring the difference $\Delta\varphi_a$ between the angles of arrival of the light rays at the two ends of the baseline. As $r_s\to 2M$, while $d_\perp(\ell)$ might be running from negative infinity to positive infinity as $\ell$ goes from $0$ to $L=2 \sqrt{r_s^2-r_{\rm min}^2(b)}$ in a window of $\theta_a(b)$, the angular difference $\Delta\varphi_a \propto 1/d_\perp$ is virtually zero for almost all $\ell$ except those of the rays coming from the region giving $d_\perp \approx 0$, where (\ref{defdphi}) breaks down [Figure \ref{anomaly} (right)]. 
In fact, what is happening in a normal case is similar: $\Delta\varphi_a \propto 1/d_\perp$ is almost a constant for most of $\ell$.
Thus, in both cases, as $r_s$ goes to $2M$, the far observer eventually cannot resolve the depth $\ell$ of the emitters from $d_\perp$ except those emitters located around the shell center $C$. 

For a near observer, the above descriptions are still valid around each layer of images with respect to the affine distance $d_A$ (e.g., Figure \ref{DisAinPS}). Unlike $d_\parallel$, however, the observed order of the layers in $d_\perp$ for $\theta_a \in (\theta_a^c, \pi-\theta_a^c)$ may be different from the order in $d_A$, and different layers may intersect, due to the presence of the negative-$d_\perp$ intervals ($P_\infty, P_0$) in the sections of light rays outside the shell. In particular, when $r_o$ is sufficiently close to $2M$ such that $\theta_1 >\pi/2$, even the light rays producing the nearest image layer with respect to $d_A$ will get $P_\infty$ outside the shell for $\theta_a \in (\pi/2, \theta_1)$, where the distance $d_\perp$ of the front surface in the nearest layer can thus be negative. In spite of this, as $r_s\to 2M$, the images of the emitters inside the shell but not around the shell center will still squeeze around the merged images of the front and rear surfaces of the shell in each layer, and the depths of those interior emitters are not resolvable by the observer from the binocular distance $d_\perp$.

\subsection{Baselines in other directions and trinocular distance}

For an observer with the baseline not exactly parallel or perpendicular to the polar direction, the tangent lines of the rays arriving at the two ends of the baseline would not intersect in general. In this case, the observer could simply rotate the baseline to where $d_\parallel$ and $d_\perp$ can be defined. As we learned earlier (e.g., Figure \ref{DisPrlx}), the value of $d_\perp$ is generally different from $d_\parallel$ of the same emitter. Since there is no rule to judge which one is better, 
the observer could further average these two distances to get a trinocular distance such as \cite{HP10} 
\begin{equation}
    d_3 \equiv  \frac{2}{(1/d_\parallel)+(1/d_\perp) }. 
\end{equation}
In Figure \ref{DisPrlx} one can compare the trinocular distance $d_3$ (green dotted) with the binocular distances $d_\perp$ and $d_\parallel$, as well as the affine distance $d_A$ of the emitter for the observer. As the emitter $e$ is brought away from the observer $O$ along the black path in Figure \ref{plotDefdphi}, one can see that the trinocular distance $d_3$ can be very close to the affine distance $d_A$ before $d_\perp$ approaches its first divergence at $P_\infty$. After $P_\infty$, the behavior of the trinocular distance $d_3$ becomes totally different from $d_A$. It diverges wherever $d_\perp=-d_\parallel$ and vanishes wherever $d_\perp$ or $d_\parallel$ vanishes.

In Figure \ref{emitterarray} we show how a spherical shell with an array of the emitters inside would be perceived in terms of $d_3$, $d_\perp$, and $d_\parallel$ by a far observer. For small angles of arrival $\theta_a$, the direct image of the front surface of the shell in terms of $d_3$ is much closer to the one in the affine distance $d_A$ [Figure \ref{emitterarray} (upper-right)] than the images in $d_\parallel$ and $d_\perp$ are. In the region in which $\theta_a$ is small and far enough from the direction of the first divergence of $d_\perp$, the properties of the images in terms of $d_3$ are similar to $d_\parallel$ and $d_\perp$,  and the spherical shell with interior emitters would look inside out for the observer. As $r_s\to 2M$, almost all the variations of the perceived distances $d_\perp$, $d_\parallel$, and $d_3$ in the depth $\ell$ between the front and rear surfaces of the spherical shell go to zero, and so the depth information of the interior emitters is not resolvable from these distances except those emitters around the shell center. A light ray emitted by a point source in this core region around the center can get a sufficiently low impact parameter $|b| < b_c$ whenever $r_s>2M$ [see Appendix \ref{EscCone}]. This allows it to easily go beyond the potential barrier peaked at the photon sphere (proportional to $b^2$ in (\ref{Eshell})) to reach the outside world even when the shell radius $r_s$ is sufficiently close to the Schwarzschild radius $2M$. Nevertheless, the size of this core region would go below the scale of any emitter in the same limit $r_s\to 2M$. Without introducing quantum effect, the final signals from this region right before the star becomes a black hole may not contain any structure interesting at the usual scale for the localized observer sufficiently far from the shell surface, though the range of the binocular distance $d_\perp$, $d_\parallel$, or $d_3$ of this region can be large when perceived by the observer.


\section{Other measures of distance}
\label{lumin}

So far, we have seen that a nearly black spherical shell would be perceived like a membrane rather than a ball, with the information of the event/emitter depths inside the shell very hard to resolve in terms of the affine and binocular distances by a localized observer outside the shell. One may wonder if there is some way to extract the depth information of the interior emitters or events more easily as the nearly black shell is not truly black. It turns out that, at least, the radar distance and the luminosity distance can work. 

\subsection{Radar distance}
\label{dradar}

Suppose the observer $O$ sends a radar signal into the spherical shell at some moment, assuming the energy input by the radar signal would not turn the nearly black star into a black hole. The wavelength of the radar signal will be blue-shifted as it drops into the shell. Suppose the mass of the pointlike object to be observed inside the shell is much greater than the energy of the blue-shifted photons of the radar signal, and the energy loss  of the radar signal in the scattering event $e$ is negligible. Then, the echoes climbing out the gravitational potential would be approximately at the same wavelength as the original radar signal when received by the observer. 

To obtain the radar distance of an event inside the shell, we start with Eq. (\ref{Eqt}), which implies $\partial_\lambda \left( A(r) \dot{t} ~\right) =0$ and so 
\begin{equation}
  \dot{t} = \frac{a}{A(r)} \label{dott}
\end{equation}
with a constant of integration $a$. The value of $a$ cannot be chosen freely because of the constraint (\ref{NullCond}). Inserting (\ref{dott}), (\ref{dottheta}), and (\ref{ETV}) into (\ref{NullCond}), one can see that $a=\sqrt{2E}=\sqrt{A_o}$ in our normalization for the affine parameter $\lambda$. Thus, for our localized observer at $r=r_o$, the radar distance of a scattering event $e$ is half of the duration from emitting to receiving the radar signals in the localized observer's proper time ($c=1$) \cite{DIn92, DG01, lin19a}, namely, with $\Delta \tau_o = \sqrt{A_o}~\Delta t$,
\begin{equation}
    d^{}_R = \frac{\Delta \tau_o}{2} = \left| \sqrt{A_o} \int_O^e \frac{a}{A(r)} d\lambda \right| 
		= A_o \int_O^e \sqrt{\frac{B(r)}{2A(r)(E-V(r))}}dr,
\end{equation}
from (\ref{ETV}) [cf. (\ref{affdist})]. Here, $\int_O^e dr= \int_{r_e}^{r_o} dr$ for $|eF| \le |eR|$, and $\int_O^e dr= (\int_{r_{\rm min}}^{r_e} + \int_{r_{\rm min}}^{r_o})dr$ for $|eF| > |eR|$ in Figure \ref{plotDefdphi} with $r_{\rm min}$ given in (\ref{rminb}). The inside-shell section of the null geodesic of the radar signal contributes
\begin{equation}
   d^{\,\rm in}_R= A_o \int_F^e dr \sqrt{\frac{1}{2A_s(E-V(r))}} 
	=\frac{A_o}{\sqrt{A_s}} \int_F^e dr \sqrt{\frac{1}{A_o- (A_s b^2/r^2)}} = \ell \sqrt{\frac{A_o}{A_s}}\label{dRinell}
\end{equation}
with the depth $\ell$ defined earlier in (\ref{affdistIn}) and (\ref{defellL}). 
The echoes from two events at different $\ell$ along the same null geodesic inside the shell could be easily distinguished by the observer's clock even if the shell radius is close to the Schwarzschild radius $2M$, since the perceived time interval $\Delta d^{\,\rm in}_R/c = \sqrt{A_o/(c A_s)}\Delta\ell$ will be dilated by a factor $1/\sqrt{A_s}$ and become significant for the observer as $A_s\to 0$. 
Therefore, the depth information of an event inside the nearly black star is in principle resolvable if the observer describes it in terms of the radar distance $d_R$, together with the angle of arrival $\tilde{\theta}_a$ of the echo of the radar signal.

\subsection{Luminosity distance}

If the emitters are some standard candles well known or even sent by the observer into the shell, then the apparent luminosity of the emitters can also reveal the information of their depths inside the shell to the observer. 

The luminosity $l$ of an isotropic emitter observed by an infinitesimal antenna of area $d{\cal A}$ is proportional to the solid angle $d\Omega$ that all the emitted light rays hitting the antenna pointing to some angle of arrival went through earlier on the unit sphere surrounding the emitter. In (3+1)D Minkowski space, $d{\cal A} = r_L^2 d\Omega$, where $r^{}_L$ is the distance from the emitter to the localized observer. For an emitter $e$ as the standard candle of unit radiated power inside the spherical shell, the apparent luminosity that the radiated power experienced by an antenna of the observer $O$ at $r_o$ outside the shell would be $(\sqrt{A_s/A_o}~)^2 \times l$ due to gravitational redshift \cite{We72}. Thus, the luminosity distance $d^{}_L$ of the emitter for the observer $O$ would be 
\begin{equation}
  d^{}_L = \sqrt{\frac{d{\cal A}}{(\sqrt{A_s/A_o}~)^2 d\Omega}} 
	= \sqrt{\frac{A_o}{A_s}
	\left|\frac{r^{}_o\theta'_o \times r_o \sin \theta_o\,d\varphi}{\theta'_d \sin \theta_d \, d\varphi}\right| }
\end{equation}
(see Figure \ref{plotDefdphi}). From (\ref{defthetad}), (\ref{thetaoIn}), and (\ref{thetaoprimeIn}), one finds that the factors 
\begin{equation}
  \frac{\theta'_o}{\theta'_d} = \left.\frac{\theta'_o}{b'}\right|_F\sqrt{\frac{A_o}{A_s}}\left(\ell-\frac{L}{2}\right)+ \frac{2}{L}\ell
\end{equation}
and
\begin{equation}
  \frac{\sin \theta_o}{\sin \theta_d} = \frac{1}{b}\sqrt{\frac{A_o}{A_s}}\left(\ell-\frac{L}{2}\right)
	  \sin \left(-\theta_{\rm in} + \left.\theta_o\right|_F\right) + \cos \left(-\theta_{\rm in} + \left.\theta_o\right|_F\right)
\end{equation}
are both linear in the depth $\ell$, and so the luminosity distance $d_L(\ell)$ is nearly linear in $\ell$ in a large portion of its range. 
The contrast between different values of $\ell$ will be enhanced by a factor of $A_o/A_s$ and so the relative depths the standard candles could be resolved more easily by the observer as $r_s\to 2M$, though the overall apparent luminosity $(\sqrt{A_s/A_o}~)^2\times l$ goes to zero in this limit.

\section{Summary and Discussion}
\label{SumDisc}

To see how a nearly black star would be perceived by a localized observer outside the star, we have investigated a simple model of a semi-transparent spherical shell with a few pointlike light emitters distributed inside. 
In Sections \ref{SpheMassShell} and \ref{binoculardistances}, we found that, in terms of the affine distance and the binocular distance determined by the outgoing light rays from the interior emitters, a localized observer outside the shell would perceive that almost all of the images of the interior emitters squeeze around the shell surfaces, while the distance between the images of the front and rear surfaces goes to zero as the shell radius $r_s$ in bookkeeper coordinates is approaching the Schwarzschild radius $2M$. 
The resolution of depth from the star surface for each interior emitter is decreasing as the star is turning to a black hole, while the whole star is perceived more and more like a 2D membrane rather than a 3D ball for the observers outside. 

For the binocular distance, this phenomenon is a consequence of the fact that, as $r_s\to 2M$, only the light rays going nearly in the radial and opposite directions from the inside of the spherical shell can go far enough to reach an observer localized outside the shell (Section \ref{dirimg} and Appendix \ref{EscCone}). Such a behavior of null geodesics in this limit is not restricted in the specific mass distribution of our model. 

Nevertheless, using alternative measures such as the radar distance or the luminosity distance, the depth information of the interior emitters is in principle resolvable when the star is not exactly a black hole, as shown in Section \ref{lumin}.

\subsection{Full knowledge about the signal}

One may be tempted to conclude that the radar and luminosity distances simply do better than the binocular and affine distances, and the entropy of the nearly black star simply jumps from a volume-scaling quantity to some area-scaling quantity with a totally different nature when the star becomes a black hole. Such conclusion might be somewhat naive, anyway. Recall that, to determine the radar distance from the observer to an event, the event must have both the past and future causal connections with the observer, and the observer must refer to the full knowledge of each received signal (echo), such as the emission time and other identities of the original radar signals, input earlier by herself. To determine the luminosity distance precisely, the observer has to know the properties of the standard candles. This cannot be fully achieved unless the standard candles were sent into the shell earlier by the observer herself or by other agents who can exchange the information with the observer. In other words, the standard candles must have two-way causal connections with the observer, too.
Actually, if the emitters inside the shell are smart enough to measure the relative locations of each other by themselves and report to the observer, then the observer certainly will know the full information of depth of the interior of the star -- if she can receive and {\it understand} the report. This again requires two-way causal connections earlier between the smart emitters and the observer. In contrast, 
the full knowledge about a received signal is not needed in determining the binocular or affine distance
(some physical assumptions on the signal source inside the shell would still be needed, though).
Only the future causal connection from the emitting event to the receiving observer is sufficient.
Thus, the observer's full knowledge about the signal coming out of a nearly black star, or the autocorrelation between the ingoing and outgoing signals, is crucial in estimating the number of degrees of freedom of the star for the outside observer. 
If an observer has the knowledge about the probes sent into a nearly black star earlier, by including or ignoring it, the observer could switch her estimates of the field degrees of freedom in the star between a volume law and a quasi-area law when analyzing the outgoing signals from the probes.

This may be related to recent observations in black hole thermodynamics: if an observer far from a black hole has a sufficiently long memory to see the autocorrelations between the Hawking radiation and the vacuum fluctuations before the black hole formed or after the black hole evaporates, then the observer will find the Hawking radiation non-thermal and the effective temperature would not be exactly proportional to the surface gravity of the black hole \cite{HSU15, Vi15}.

\subsection{Area law at late times of gravitational collapse}

When the collapsing star is so close to a black hole that each signal from its interior is too weak or too red-shifted to be detected or resolved by any observer outside, all those distances to the interior emitters or events will not be measurable in the outside world. 
Before this happens, however, the radar and luminosity distances would have been harder to determine than the binocular distance 
because of the need of a very long memory. For example, a radar signal may spend a time for a round trip much longer than the observer's lifetime in her clock, and the standard candles sent by the observer may take a similar time-scale in the observer's clock to spread inside the star in order to explore the interior volume. Further, when the star is about to become a black hole, sending energy such as radar signals or standard candles into the star may turn the star to a black hole, such that the echoes of the radar signals or the light emitted by the standard candles would never reach the outside world. Thus, for the observers witnessing the last stage of black hole formation, more likely, they would perceive the star like a 2D membrane rather than a 3D ball according to the binocular distance, and the area law would eventually dominate after the thickness of the membrane is below the Planck length and not resolvable. At late times, when {\it every} observer outside the star fails to keep or learn any historical knowledge of the received signals before they are emitted or scattered from the interior, the physical entropy of the nearly black star for the outside world would follow the area law.

\subsection{Estimate of field degrees of freedom in a black hole}

For a far observer, the observational data on the interior points at the angle of arrival $\tilde{\theta}_a \ge \tilde{\theta}_1$ are repeating (all are indirect images; see Figure \ref{theta1} and Section \ref{dirimg}), and so they should not count in the number of field degrees of freedom in the star. 
When the shell radius $r_s$ is sufficiently close to the Schwarzschild radius $2M$, all the direct images observed at the angles of arrival from $\tilde{\theta}_a =0$ to $\tilde{\theta}_1$ can be mapped back to the region of the shell surface with $\theta$ between $0$ and $\pi/2$ in bookkeeper coordinates, namely, the half sphere facing the observer. Since the depth information of the points cannot be resolved, the number of the field degrees of freedom of the star would be proportional to the area of the half-sphere $A/2$ with $A = 4\pi r_s^2$.
In Ref.\cite{Be73}, Bekenstein argued that the minimum increase in area of a Kerr black hole by dropping a particle into it is $2\hbar$.  Interestingly enough, if we take $2\hbar$ as the unit area element of a black hole horizon, or the specific area of a field, then for the outside world, the field degrees of freedom in the black hole evolved from our spherical star could be estimated as $(A/2)/(2\hbar) = A/(4\hbar)$, which has the same value as the Bekenstein-Hawking entropy, although how the field degrees of freedom is proportional to entropy is not clear here yet.
 
\begin{acknowledgments}
I would like to thank Bill Unruh and Bei-Lok Hu for illuminating discussions and Chiang-Mei Chen, Hung-Yi Pu, Chopin Soo, and Yi Yang for helpful information. I also thank the anonymous referee for valuable comments and suggestions.
This work is supported by the Ministry of Science and Technology of Taiwan under Grant No. MOST 106-2112-M-018-002-MY3 
and in part by the National Center for Theoretical Sciences, Taiwan.
\end{acknowledgments}

\appendix

\section{Null geodesics in a spherically symmetric (3+1) dimensional spacetime}
\label{RevGeoEq}

Suppose a spherical star collapses radially in a very slow rate, 
so slow that in the period of our interest the spacetime geometry can be approximately described by the static, spherically symmetric metric,
\begin{equation}
  ds^2=-A(r) dt^2 + B(r) dr^2 + r^2 \left( d\theta^2 + \sin^2\theta d\varphi^2\right). \label{SphereMetric}
\end{equation} 
A light ray in this background geometry satisfies the geodesic equations
\footnote{A more efficient derivation for (\ref{NullCond})--(\ref{EAo}) directly using the symmetry of the system can be found in, e.g., Ref. \cite{HP02}.}
\begin{eqnarray}
  &&\ddot{t} + \frac{A'}{A}\dot{r}\dot{t} = 0, \label{Eqt} \\
  &&\ddot{r} +\frac{A'}{2B}\dot{t}^2+\frac{B'}{2B}\dot{r}^2-\frac{r}{B}\dot{\theta}^2-\frac{r}{B}\sin\theta\dot{\varphi}^2=0, \label{Eqr} \\ 
  &&\ddot{\theta} + \frac{2}{r}\dot{r}\dot{\theta} -\sin\theta\cos\theta \, \dot{\varphi}^2 =0, \label{Eqtheta} \\
	&&\ddot{\varphi} + \frac{2}{r}\dot{r}\dot{\varphi}+2 \cot\theta\, \dot{\varphi}\dot{\theta} =0 \label{Eqphi}
\end{eqnarray}
and the null condition
\begin{equation}
  -A \dot{t}^2 + B \dot{r}^2 + r^2 \dot{\theta}^2 +r^2 \sin^2\theta \,\dot{\varphi}^2 =0, \label{NullCond}
\end{equation}
where the dots and primes denote the derivatives with respect to some affine parameter $\lambda$ and the $r$-coordinate, 
respectively. Equation (\ref{Eqphi}) implies $\partial_\lambda( r^2 \sin^2\theta \, \dot{\varphi} )=0$,
and so along the light ray, one has 
\begin{equation}
  r^2\sin^2\theta\, \dot{\varphi} =K, \label{aziangmtm}
\end{equation} 
which is a constant interpreted as the effective (with respect to the affine parameter $\lambda$ rather than proper time) azimuthal angular momentum ``per unit mass" of the light.
If $K$ is not zero, we can insert $\dot{\varphi} = K/(r^2 \sin^2\theta)$ into Eq. (\ref{Eqtheta}) and then multiply the equation by $\dot{\theta}$ to obtain $\partial_\lambda\{r^2[(r\dot{\theta})^2 + (r\sin\theta\,\dot{\varphi})^2]\}=\partial_\lambda [ r^4 \dot{\theta}^2 + (K^2/\sin^2\theta)]=0$, or
\begin{equation}
    r^4 \dot{\theta}^2 = J^2 - \frac{K^2}{\sin^2\theta}, \label{EqthetaDotSq}
\end{equation}
where $J$ is another constant of motion, interpreted as the effective total angular momentum of the light.

Suppose an observer is localized at some point outside the star, and we choose the $z$ axis joining the center of the star [the origin in bookkeeper coordinates \cite{TW00} given in (\ref{SphereMetric})] and the localized observer. Then, in bookkeeper coordinates the observer would be localized around a point of $\sin\theta=0$ while $r\not=0$. Since we are looking into how the star light would be perceived by the observer, we are only interested in the light rays passing through the localized observer. However, for any finite $J^2$ and nonzero $K$, the right-hand side of (\ref{EqthetaDotSq}) diverges to negative infinity as $\sin\theta \to 0$, while the left-hand side is positive definite. Thus, $K$ has to be zero for the light rays seen by the observer, and these light rays must have $\dot{\varphi}=0$ off the $z$ axis from (\ref{aziangmtm}). In other words, each light ray from a pointlike emitter to the localized observer on the $z$ axis will be lying on a constant-$\varphi$ hypersurface by symmetry.

Allowing that $\theta$ can be negative while requiring $\dot{\varphi} = 0$ in Eq. (\ref{Eqtheta}), we find $\partial_\lambda (r^2 \dot{\theta} ) = 0$, and so
\begin{equation}
  r^2 \dot{\theta} = b
	\label{dottheta}
\end{equation}
with a constant $b$, interpreted as the effective polar angular momentum of the light.
Eliminating $\dot{t}^2$ by (\ref{NullCond}) and then introducing (\ref{dottheta}), Eq. (\ref{Eqr}) becomes
\begin{equation}
  \partial_\Lambda \partial_\Lambda r = -V'(r), \label{effEOM}
\end{equation}
where $\partial_\Lambda \equiv \sqrt{AB}\partial_\lambda$ can be thought of as the effective time-derivative operator and 
\begin{equation}
  V(r) = \frac{b^2 A(r)}{2r^2}  \label{Veff}
\end{equation}
can be thought of as the effective potential for radial motion of a particle of unit mass. Multiplying both sides of (\ref{effEOM}) by $\partial_\Lambda r$, one finds $\partial_\Lambda [ \frac{1}{2}\left( \partial_\Lambda r\right)^2 + V ] = 0$, and thus
\begin{equation}
  \frac{1}{2}\left( \sqrt{A(r)B(r)}\,\dot{r}\right)^2 + V(r) = E \label{ETV}
\end{equation}
with a constant of motion $E$, interpreted as the effective total energy of that particle.\footnote{
Conventionally one writes $E=({\cal E}/m)^2/2$ for the geodesic of a massive particle of energy per unit mass ${\cal E}/m$, which goes to $1$ as $r_o\to \infty$ \cite{TW00}.}

Denote the radius of the star by $r_s$ and the position of the localized observer 
$O$ by $(r_o,\theta_o)$ with $r_o>r_s$ and $\theta_o=0$ on the $r\theta$ plane of constant $\varphi$. 
To match the affine parameter $\lambda$ to the local radar distance (cf. Section \ref{dradar}) 
around the observer determined by the observer's proper time $d\tau_o^2 = A(r_o)dt^2$, 
we impose the normalization condition 
\begin{equation}
  \left. \dot{r}^2 \right|_{r=r_o} = \frac{1}{B(r_o)}\left(1 - r_o^2 \dot{\theta}^2(r_o) \right) \label{BCrdot}
\end{equation}
from (\ref{NullCond}) for $\lambda$. In this normalization (\ref{ETV}) is simply
\begin{equation} 
  E=\frac{A(r_o)}{2}\equiv \frac{A_o}{2}  \label{EAo}
\end{equation} 
after (\ref{dottheta}) is inserted. Given the metric components $A(r)$ and $B(r)$, and the position of the localized observer $r_o$, the only free parameter for the null geodesics of our interest is $b$. 

In practice, one obtains the light ray from the event/emitter $e$ at ($r_e, \theta_e$) to the observer $O$ on the $r\theta$ plane in bookkeeper coordinates by first solving $r(\lambda)$ from (\ref{ETV}) with (\ref{EAo}) and the initial conditions (\ref{BCrdot}) and $r=r_o$ at $\lambda=0$, then integrating (\ref{dottheta}) to get 
\begin{equation}
   \theta_e - \theta_o = \int_{\theta_o}^{\theta_e} d\theta = \int_O^e \frac{b}{r^2}d\lambda = 
		\int_O^e \frac{b}{r^2} \sqrt{\frac{A(r)B(r)}{2(E-V(r))}}dr . \label{thetaofr0}
\end{equation}
Here $\int_O^e$ denotes that the integration should be done from $O$ to $e$ along the null geodesic in the direction of increasing $\lambda$. According to (\ref{dottheta}), $\theta_e - \theta_o$ should be positive when $b>0$. For example, if $e$ is outside the spherical shell in Section \ref{SpheMassShell} and $O$ is outside the photon sphere ($r_o>3M$; see Figure \ref{Vofr}), then $r$ will be a single-valued function of $\lambda$. Inserting (\ref{Schsch}), (\ref{Eshell}), and (\ref{EAo}) into (\ref{thetaofr0}), and denoting $r^{}_< ={\rm min}\{r_o,r_e\}$ and $r^{}_> ={\rm max}\{r_o,r_e\}$, one has 
\begin{equation}
   \theta_e - \theta_o = \int_{r^{}_<}^{r^{}_>} \frac{b\,d r}{\sqrt{A_o r^4-b^2 r^2+2Mb^2 r}},
   \label{thetaofr}
\end{equation}
which is an elliptic integral and can easily be calculated numerically. 
For the cases with both the emitter $e$ and the observer $O$ being inside the photon sphere ($2M < r_s<r_o < 3M$), the $r$ integration in (\ref{thetaofr0}) may not be that simple since $r(\lambda)$ could oscillate in an interval $(r_{\rm min}, r_{\rm max})$ containing $r_e$ and $r_o$. In these cases, the $r$ integration in (\ref{thetaofr0}) would be done piecewise over sub-domains in each of which $r(\lambda)$ is single-valued, and one should take care of the sign in each piece of the $r$ integration to make the result monotonic in $\lambda$.

Suppose the observer $O$ is very far from the shell ($r_o\gg 2M$), then $A_o\approx 1$. 
For $r \gg b$, (\ref{thetaofr}) implies $\theta(r) - 0 \approx \int_{r}^\infty b dr'/ r'^2 = b/r$ asymptotically, such that the constant $b \approx r \theta \approx r \sin\theta  = \rho$ is the distance from the point $(r,\theta, \varphi)$ on the light ray to the $z$ axis joining the shell center $C$ and the localized observer $O$ in bookkeeper coordinates. Thus, $b$ can be interpreted as the impact parameter for the incident photons from $r\to \infty$.

\section{Refraction and escape cones}
\label{EscCone}

When a light ray departs from a pointlike emitter situated at radius $r_e < r_s$ in the spherical shell, the angle of departure $\theta_d$ about the radial direction of the shell in bookkeeper coordinates (see Figure \ref{plotDefdphi}) is given by
\begin{equation}
  \tan \theta_d = \left.\frac{r\dot{\theta}}{\dot{r}} \, \right|_{r\to r_e}.
\label{defthetad}
\end{equation}
Later, when the light ray is crossing the shell surface from the inside to the outside of the spherical shell, the angle of incidence $\theta_{\rm in}$ and the angle of transmission $\theta_{\rm out}$ around the shell surface on the $r\theta$ plane are given by
\begin{equation}
  \tan \theta_{\rm in} = \left. \frac{r\dot{\theta}}{\dot{r}}\,\right|_{r\to r_s-}, \hspace{1cm} 
  \tan \theta_{\rm out} = \left. \frac{r\dot{\theta}}{\dot{r}}\,\right|_{r\to r_s+}.
	\label{thIthT}
\end{equation}
Since $r\dot{\theta}=b/r$ from (\ref{dottheta}) is continuous around $r=r_s$, one may arrange (\ref{thIthT}) into the form of Snell's law of refraction, $n_g(r_s-\epsilon) \sin \theta_{\rm in} = n_g(r_s+\epsilon)\sin\theta_{\rm out} = r \dot{\theta}|_{r=r_s} = b/r_s$, where $\epsilon\to 0+$ and the effective index of refraction on the $r\theta$ plane is defined as 
\begin{equation}
   n_g(R) \equiv \left.\sqrt{\dot{r}^2 + r^2 \dot{\theta}^2} \right|_{r=R}.
\end{equation}
It is obvious that $n_g(\infty)=1$ for the observer at $r_o\to \infty$ from (\ref{dottheta}), (\ref{ETV}), and (\ref{Eshell}),
which also yield
\begin{eqnarray} 
  \theta_d &=& \tan^{-1}\left[\left(\frac{L}{2}-\ell \right)\frac{n_g^-}{b}\right] - {\rm sgn}(b) \frac{\pi}{2}
	\label{thetad}\\
	\theta_{\rm in} &=& -\sin^{-1}\frac{b}{r_s n_g^-}, \hspace{.5cm} \theta_{\rm out} = -\sin^{-1}\frac{b}{r_s n_g^+}, \label{thetaio}
\end{eqnarray}
with the depths $\ell(b)$ and $L(b)$ defined in (\ref{defellL}), and the effective indices of refraction $n_g^-\equiv n_g(r_s-\epsilon) =\sqrt{2E/A_s}$ and $n_g^+ \equiv n_g(r_s+\epsilon)=\sqrt{2E+ 2M b^2/r_s^3}$. 
We have an overall minus sign in each angle in (\ref{thetaio}) because in our parametrization the affine parameter $\lambda$ is monotonically decreasing along the null geodesic started at the emitter $e$ (recall $\lambda\equiv 0$ at $O$). 
Without considering any specific observer, $E$ does not have to be $A_o/2$ here but is a free parameter satisfying the condition 
$E\ge V(r_e)= b^2 A_s/(2 r_e^2)$ from (\ref{ETV}). Note that $n_g^+$ here depends on $b$, as $\theta_{\rm in}$, $\theta_{\rm out}$, and $\theta_d$ all do. Although $n_g^+/n_g^- = \sqrt{A_s(1 + Mb^2/(E r_s^2))}< 1$ for all $r_{\rm min}<r_s$ and $r_s>2M$ from (\ref{rminb}), there is no total internal reflection around the shell surface. Indeed, $\theta_{\rm out} = \sin^{-1}\{(2E/b^2)r_s^2+1-A_s\}^{-1/2} \le \sin^{-1}\{A_s(r_s^2-r_e^2)/r_e^2 +1 \}^{-1/2} < \pi/2$ whenever $A_s >0$ and $r_e < r_s$.

When the spherical shell is gravitationally intense ($2M < r_s < 3M$), only the light rays emitted with $E\ge V(3M)=b^2/(54M^2)$ can cross the barrier of the effective potential $V$ around the photon sphere $r=3M$ and escape to future null infinity. These light rays must have their impact parameters $b$ less than $b_c$ given in (\ref{bcrit}), and so their $|\theta_d|$, $|\theta_{\rm in}|$, and $|\theta_{\rm out}|$ will not exceed 
\begin{eqnarray}
  \theta_d^c &\equiv& \tan^{-1} \left[ \frac{r_e^2}{27 M^2 A_s}-1\right]^{-1/2} = 
	\sin^{-1} \frac{3\sqrt{3A_s}M}{r_e},  \label{thetadc} \\
  \theta_{\rm in}^c &\equiv&\sin^{-1} \frac{3\sqrt{3A_s}M}{r_s},  \hspace{.5cm}
  \theta_{\rm out}^c \equiv \tan^{-1} \left[\frac{r_s^2}{27M^2}-A_s\right]^{-1/2},
\end{eqnarray}
respectively. An interior emitter at $r_e > 3\sqrt{3A_s}M$ has two ``escape cones" \cite{Pe04, Sy66, Ja70} in the $\pm r$ directions bounded by $\theta=\theta_d^c$ and $\pi-\theta_d^c$, respectively, in bookkeeper coordinates. For an emitter situated in the vicinity of the shell center with $r_e \le 3\sqrt{3A_s}M$, Eq. (\ref{thetadc}) breaks down, and the light rays sourced from this emitter can escape to null infinity in all directions since $E\ge V(r_e)= b^2 A_s/(2 r_e^2) \ge b^2 A_s/(2(3\sqrt{3A_s}M)^2)=V(3M)$ for all $b$ here. As $r_s \to 2M$, both $\theta_d^c$ and $\theta_{\rm in}^c$ go to zero for $r_e > 3\sqrt{3A_s}M \to 0$. Thus, for almost all the interior emitters but those at the shell center, only the light rays emitted in the radial and opposite directions in bookkeeper coordinates can escape the photon sphere and reach a far observer localized at $r_o > 3M$.

For a near observer inside the photon sphere ($2M < r_s < r_o < 3M$), the situation is similar. Consider the light rays started at an emitter at $r=r_e$ with the angle of departure $\theta_d$. Rewrite (\ref{thetad}) as 
\begin{equation}
  b = -\sqrt{\frac{2E}{A_s}} \,r_e  \sin \theta_d.  \label{bofthetad}
\end{equation}
Then, the maximal $r$ that the light ray can reach, $r_{\rm max}$, satisfies $V(r_{\rm max})=E$ from (\ref{ETV}), or
\begin{equation}
   A_s r_{\rm max}^3 = r_e^2 \sin^2 \theta_d \left(r_{\rm max}-2M\right)  \label{rmaxcond}
\end{equation}
after (\ref{Veff}) is inserted. Given a fixed value of $r_o \in (2M, 3M)$, (\ref{rmaxcond}) implies that the light rays of $\theta_d$ significantly deviate from $0$ or $\pi$ (i.e., $\sin^2\theta_d$ is not too small) from an interior emitter off the shell center (i.e., $r_e$ is not too small) can never be seen by the observer (i.e., $2M\alt r_{\rm max} < r_o$) when $r_s$ is sufficiently close to $2M$ (i.e., $A_s\to 0$ and the left-hand side of (\ref{rmaxcond}) is negligible).
$\theta_{\rm in}$ of the same light ray is squeezed to zero in the same limit since $|\theta_{\rm in}|=|(r_e/r_s)\theta_d| < |\theta_d|$ from (\ref{thetaio}). These ensure that once a localized observer at $r=r_o$ is not too close to the shell surface $r_s$, even she is inside the photo sphere, the inside-shell section(s) of each light ray observed by her must have gone almost along some diameter(s) passing through the center of the nearly black star. 

Therefore, when the spherical shell is about to form a black hole, the distance-determination schemes using different view angles of objects do not work for an outside observer looking at the pointlike emitters or events inside the shell (except those emitters at the shell center). A localized observer outside the shell cannot determine the relative depths of two emitters by comparing the difference of the relative positions of their direct images and those of their indirect images, and the binocular distances of almost all the emitters along a diameter of the star passing through the shell center become indistinguishable for the localized observer. 
In terms of the binocular distances, the direct images of the emitters inside the nearly black star would be seen by different observers outside the shell as the same pattern distributed in a 2D membrane up to a translation of the image center of the whole star together with the periodic boundary. Each observed pattern can be mapped back to the half-sphere of the shell facing the observer.

\end{document}